\documentclass[aps,prb,reprint,groupedaddress]{revtex4-2}

\usepackage[utf8]{inputenc}
\usepackage[english]{babel}

\usepackage{microtype}
\DisableLigatures[f]{encoding = *, family = *}

\usepackage{hyperref}
\hypersetup{colorlinks=true, linkcolor=[rgb]{0 0 0.6}, citecolor=[rgb]{0 0 1}, urlcolor=[rgb]{0 0 1}}

\usepackage{enumitem}
\usepackage{amssymb}
\usepackage{amsfonts}
\usepackage{mathrsfs}
\usepackage{empheq}
\usepackage{tikz-cd}
\usepackage{amsmath}
\usepackage{graphicx}
\usepackage{dsfont}
\usepackage{dcolumn}
\usepackage{bm}
\usepackage{color,soul}
\usepackage{xcolor}
\usepackage{url}
\usepackage{tabularx}
\usepackage{array}
\usepackage{booktabs}
\usepackage{comment}
\usepackage{hyperref}
\usepackage{footnote}
\usepackage{lipsum}
\usepackage{mathrsfs} 
\usepackage[normalem]{ulem}

\definecolor{GREEN}{rgb}{0.,0.5,0}

\begin{document}

\title{
Topological invariants of vortices, merons, skyrmions, and their combinations in continuous and discrete systems}

\author{Filipp N. Rybakov}
\email[]{philipp.rybakov@physics.uu.se}
\affiliation{Department of Physics and Astronomy, Uppsala University, Box-516, Uppsala SE-751 20, Sweden}

\author{Olle Eriksson}
\affiliation{Department of Physics and Astronomy, Uppsala University, Box-516, Uppsala SE-751 20, Sweden}
\affiliation{Wallenberg Initiative Materials Science, WISE, Uppsala University, Box-516, Uppsala SE-751 20, Sweden}

\author{Nikolai S. Kiselev}
\affiliation{Peter Gr\"unberg Institute, Forschungszentrum J\"ulich and JARA, 52425 J\"ulich, Germany}

\date{\today}

\begin{abstract}
Magnetic vortices and skyrmions are typically characterized by distinct topological invariants. 
This work presents a unified approach for the topological classification of these textures, encompassing isolated objects and configurations where skyrmions and vortices coexist.
Using homotopy group analysis, we derive topological invariants that form the free abelian group, $\mathbb{Z}\times\mathbb{Z}$.
We provide an explicit method for calculating the corresponding integer indices in continuous and discrete systems. 
This unified classification framework extends beyond magnetism and is applicable to physical systems in general. 
\end{abstract}

\maketitle

\section{Introduction}

Similar to domain walls~\cite{MalozemoffSlonczewski_1979, Hubert}, vortices represent fundamental magnetic textures~\cite{DeSimone_chapter4_2006, BraunReview2012, TopologyMagnetism2018}.
By vortices, we refer to a broad class of microscopic states characterized by curling magnetization. 
Examples include rectangular Landau patterns~\cite{LL1935, PhysRevB.72.214409, Arrott2011, fromLPatterntoVortex2022}, 
circular textures commonly observed in nanodisks~\cite{Shinjo2000, Raabe2000, Guslienko_2001, HOLLINGER2003178, Komineas_2006}, 
and swirl-like structures that form constituent parts of more complex magnetic textures~\cite{HuberSmithGoodenough1958, Feldtkeller1965, Nakatani_1989, PhysRevLett.99.117202, Pereiro2014, PhysRevB.91.224407, Yu2018, PhysRevB.101.064408}.
Moreover, magnetic textures in annular rings~\cite{Zhu2000, PhysRevLett.86.1102, Klaui_2003, MuratovOsipov2009} can also be associated with vortices.
The latter textures are often interpreted as coreless vortices, with the absent core assumed to be virtually present within the cavity.

Magnetic vortices can emerge due to various factors, including sample geometry, demagnetizing fields, and curvature effects~\cite{PhysRevB.58.9217, Moser2004, DeSimone_chapter4_2006, PhysRevB.85.144433, Streubel_2016, Volkov2024}.
Additionally, vortices can form in systems with magnetocrystalline anisotropy or, more generally, spin-orientation anisotropy.
In particular, in systems where strong hard-axis (easy-plane) anisotropy effectively reduces the spin degree of freedom from a sphere to a circle, Berezinskii-Kosterlitz-Thouless (BKT) vortices emerge~\cite{Berezinskii_1972, Kosterlitz_1973, FourtyYearsBKT, SvistunovBabaevProkofev}.
We start in fact the discussion of a unified approach for the topological classification of magnetic textures by considering a model similar to the BKT model. 
After that we soften the constraint in the spin space and consider a more general case where the spins in the vortex domain can have any direction. 
Such vortices are commonly referred to as merons~\cite{Senthil_2004, Pereiro2014, PhysRevB.91.224407, Yu2018, GOBEL20211, Augustin2021, Strungaru2022}, a term that, in a sense, denotes fractional configurations~\cite{ALLAN1977375, GROSS1978439, Volovik_book}. 
It is worth noting that nowadays the term ``meron'' is often interchangeably used with the term ``vortex''~\cite{VanWaeyenberge2006, PhysRevB.75.012408, PhysRevLett.99.117202, PhysRevB.101.064408}, making these two terms synonymous in many practical contexts. 
A key aspect of merons is that they may become constituent parts of other topological objects. 
For instance, certain combinations of merons can form solitons associated with the homotopy group~$\pi_2(\mathbb{S}^2)=\mathbb{Z}$~\cite{Hatcher} and are known as skyrmions. 
Such meron pairs are often referred to using distinct terms, such as 
bimeron~\cite{Zhang2015, PhysRevB.99.060407, Gao2019, GOBEL20211, PhysRevB.108.014402}, 
bounded half-lumps~\cite{PhysRevD.82.125030, PhysRevD.87.125013}, 
or in-plane skyrmions~\cite{moon2019}, 
but sometimes they are simply called ``vortices'' and ``antivortices''~\cite{PhysRevB.75.012408, PhysRevLett.99.117202}. 
Meanwhile, many $\pi_2(\mathbb{S}^2)$~skyrmions do not consist of merons~\cite{Thiele_1969, Bobeck_Scovil_1971, Hassan2024, KovalevKosevichMaslov, KOSEVICH1990117, Bogdanov_89, Kiselev_2011, Nagaosa2013, Melcher_2014, PhysRevB.99.064437, PhysRevB.102.144422, Bernand-Mantel2021, KIRAKOSYAN2006413, Leonov2015, PhysRevB.95.094423, Bogolubskaya1990, Piette1995, Ward1995, Manton_Sutcliffe_2004}. 
The central result of this work demonstrates that in complex systems, where both merons and skyrmions coexist, topological invariants form the free abelian group~$\mathbb{Z}\times\mathbb{Z}$  (or~$\mathbb{Z}\oplus\mathbb{Z}$ in equivalent notation) and thus such configurations must be classified in terms of an ordered pair of integers. 
An explicit method for computing these topological invariants is provided.


\section{\label{sec_S1} Spins on a circle}

\begin{figure*}
	\includegraphics[width=14.4cm]{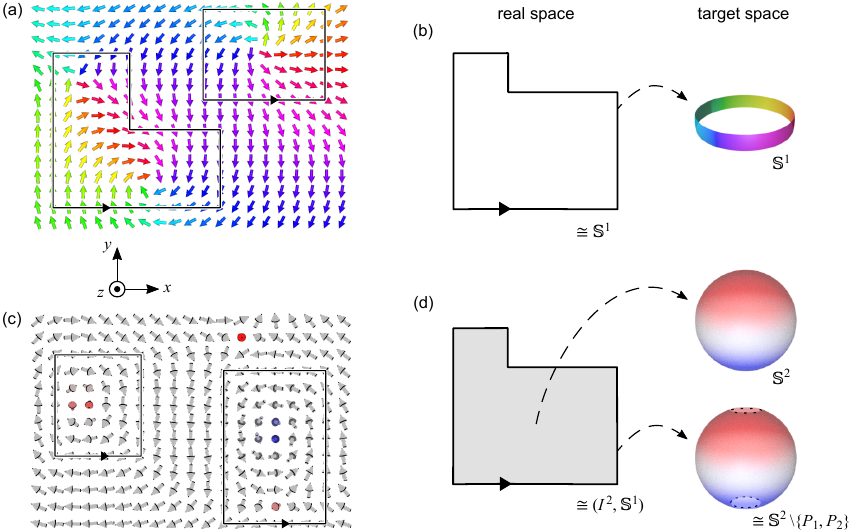}
\caption{\label{fig_latt_and_map}
Sketches of vortices, merons, skyrmions, and corresponding maps. 
(a)~Vortices in the $XY$-model.
(b)~Mapping a closed contour into spin space representing a circle. 
(c)~Merons and skyrmions in the hard axis magnet model.
(d)~Mapping a closed contour and its interior into the subspace and full space of the spin sphere, respectively.
}
\end{figure*}

In this section, we consider a system similar to the BKT model (also known as the classical $XY$-model) with spins that take values in the circle, $\mathbf{m}(\mathbf{r}_i)\in\mathbb{S}^1$.
For simplicity, but without loss of generality, let the spins be located at the nodes of a square lattice, see Fig.~\ref{fig_latt_and_map}(a).
If the dominant interactions are ferromagnetic, then it is extremely unfavorable for a pair of neighboring spins to align antiparallel. 
Accordingly, the probability of encountering antiparallel-oriented neighboring spins while traversing the lattice becomes vanishingly small.
Consequently, we restrict our consideration to closed paths governed by a simple rule, referred to as \mbox{\textit{Axiom}~I}: the angle between any two adjacent $\mathbb{S}^1$-spins along the path must never reach~$\pi$. 
Then the following continuous map takes place: 
\begin{equation}
g:\ \mathbb{S}^1\rightarrow\mathbb{S}^1 . \label{S1toS1}
\end{equation}
The circle $\mathbb{S}^1$ on the left-hand side of the map Eq.~\eqref{S1toS1} corresponds, up to homomorphism, to a simple closed curve in the real space, see Fig.~\ref{fig_latt_and_map}(b). 
The circle on the right-hand side of the map Eq.~\eqref{S1toS1} is due to a closed chain of minor arcs, formed by adjacent spins $\mathbf{m}(\mathbf{r}_{i})$ and~$\mathbf{m}(\mathbf{r}_{i+1})$.
This map is characterized by homotopy classes, distinguished by the degree only~\cite{HuHomotopyTheory, Mapping_degree_theory}. 
The associated integer invariant is given by ${w = \text{deg}(g)}$. 
Using the tangent half-angle formula, we obtain the following expression for the winding number: 
\begin{equation}
w(\mathbf{m}) = \frac{1}{2\pi}\sum\limits_{i=1}^{N}
2\arctan\left(
\frac{\hat{\mathbf{e}}_z\cdot[\mathbf{m}(\mathbf{r}_{i})\!\times\!\mathbf{m}(\mathbf{r}_{i+1})]}
{1 + \mathbf{m}(\mathbf{r}_{i})\cdot\mathbf{m}(\mathbf{r}_{i+1})}
\right),
\label{w_S1}
\end{equation}
where $\mathbf{r}_{N+1} = \mathbf{r}_1$ since we consider closed paths only. 
The winding number Eq.~\eqref{w_S1} has the following meaning. 
When ${w=1}$, the loop is said to contain a vortex. 
Such a case is illustrated by the top-right rectangular contour in Fig.~\ref{fig_latt_and_map}(a). 
When ${w=-1}$, the contour is typically described as containing an antivortex. 
States with $w=\pm 2, \pm 3, ...$ are often referred to as multivortices, vortex combinations, vortex clusters, or similar terms. 
If~${w=0}$, the contour either encloses an equal number of vortices and antivortices or no vortices at all. 
If the map in Eq.~\eqref{S1toS1} is further equipped with a base point, the homotopy classes form a group isomorphic to the integers under addition, ${\pi_1(\mathbb{S}^1) = \mathbb{Z}}$, for details see Refs.~\cite{HuHomotopyTheory, Hatcher}.

One has to emphasize that the rigorous topological arguments supporting Eq.~\eqref{w_S1} do not rely on the spatial continuity of~$\mathbf{m}$ at all.
The only essential requirement is \textit{Axiom}~I, which is naturally applicable in most practical cases, where the ferromagnetic exchange dominates other interactions.
%


\section{\label{sec_product} Functions with values in product space}

In this section, we generalize the result from the preceding section to a broader class of functions. 
We begin from a pure algebraic topology perspective. 
Specifically, consider the fundamental group of the product space: 
\begin{equation}
\pi_1(\mathbb{S}^1\times X) = \pi_1(\mathbb{S}^1) \times \pi_1(X), 
\label{prodactSxX}
\end{equation}
where $X$ is any connected space. 
By choosing $X$ to be simply connected space, \textit{i.e.}, $\pi_1(X)=0$, we naturally extend the result of the preceding section.  
Namely, we now consider closed paths following the only rule -- \mbox{\textit{Axiom~II}}: the function along the path takes values in the space homeomorphic to ${\mathbb{S}^1\times X}$, while~$X$ is simply connected and \textit{Axiom~I} holds for the subspace~$\mathbb{S}^1$. 
As a next step, we address the applicability of the \textit{Axiom}~II to physical models.

The product space in Eq.~\eqref{prodactSxX} corresponds to the general case, while here, we focus on a specific case where $X$ is either a closed or an open interval,~$I^1$. 
Consequently, the function along the path can, for example, take values in a spherical segment or in a sphere with two points removed, $\mathbb{S}^2 \setminus \{P_1, P_2\}$. 
This scenario naturally arises in classical spin models with spin-orientation anisotropy of the hard-axis (easy-plane) type.
By spin-orientation anisotropy we mean the dependence of energy on the spin direction, and this can be the result of a combination of many factors such as an applied magnetic field, magnetocrystalline anisotropy~\cite{Skomski}, shape anisotropy~\cite{GioiaJames1997, Slastikov_config_aniso, ReducedEnergies2024}, etc. 
For instance, in the case of a hard axis parallel to~$\hat{\mathbf{e}}_z$, as shown in Fig.~\ref{fig_latt_and_map}(c), sites with  $\mathbf{m}(\mathbf{r}_{i})=\pm\hat{\mathbf{e}}_z$ are highly unlikely.
The probability of the spin passing through the poles becomes vanishingly small, and thus, one can naturally accept \textit{Axiom}~II.
The winding number can then be computed using Eq.~\eqref{w_S1} for the normalized projected vectors defined as: 
\begin{equation}
\mathbf{m}\mapsto 
\frac
{\mathbf{m} - \hat{\mathbf{e}}_z (\mathbf{m}\cdot\hat{\mathbf{e}}_z)}
{|\mathbf{m} - \hat{\mathbf{e}}_z (\mathbf{m}\cdot\hat{\mathbf{e}}_z)|}.
\label{m_xy}
\end{equation}
This approach applies to both discrete and continuous systems. 
In the continuum limit, Eqs.~\eqref{w_S1} and~\eqref{m_xy} yield the well-known integral for the winding number~\cite{Mapping_degree_theory}:
\begin{equation}
w(\mathbf{m}) = \frac{1}{2 \pi}\oint d{\mathbf{l}} \cdot  
\left(
\frac{
m_x \mathbf{\nabla}m_y - m_y \mathbf{\nabla}m_x
}{m_x^2 + m_y^2}
\right). \label{winding}
\end{equation}

Remarkably, the above results extend to a wide range of physical systems beyond magnetism. 
Among these are complex scalar field models with a potential term in the form of a Mexican hat. 
In such cases, the function~$\psi$ take values in the complex plane ${\mathbb{C} = \mathbb{R}^2}$, where ${\psi = 0}$ is highly unfavorable, as it corresponds to the maximum potential energy.  
Consequently, it is natural to consider paths where the function values belong to the space ${\mathbb{R}^2 \setminus \{0\} \cong \mathbb{S}^1 \times \mathbb{R}}$. 
From this, it immediately follows that the system can host vortices characterized by the homotopy group~$\mathbb{Z}$.  
It is important to note that the above analysis is not intended to be comprehensive because some systems are more complex.
The concrete representative examples include superfluid $^3$He models~\cite{Volovik_Mineev_1976, Monastyrsky, Volovik_book, PhysRevResearch.2.023263} and multicomponent Ginzburg-Landau models~\cite{PhysRevLett.89.067001, SvistunovBabaevProkofev, PhysRevB.87.014507}.

It is important to emphasize that when \textit{Axiom}~II is not applicable, there are two possibilities: the system may lack vortices entirely, or it may exhibit vortices of a different nature.
For instance, fully isotropic ferromagnets do not support vortices~\cite{BP1975, Mineev_book, Kleman_Lavrentovich_2003}. 
In contrast, liquid crystals can host vortices associated with elements of finite groups~\cite{KlemanMichelToulouse_1977, RevModPhys.51.591, Kleman_Lavrentovich_2003, Wu_Smalyukh_review_2022}.
Finally, systems with multiple hard axes may host non-abelian vortices corresponding to elements of free groups~\cite{Rybakov_Eriksson, PhysRevResearch.6.L032011, PhysRevB.110.094442}.

\section{\label{sec_S2} Spins on a sphere}

Until now, we have focused exclusively on the spin configurations along the loop, excluding the texture inside the loops from the analysis. 
With this approach, local critical processes within the loop, such as the reversal of vortex core polarity~\cite{VanWaeyenberge2006, PhysRevLett.100.027203} or skyrmion switching/collapse~\cite{PhysRevB.79.224429, heo2016switching, Bessarab2018, Muckel2021, MicromagneticLifetime2022}, do not alter the winding number. 
In this regard, it seems important to also account for the states in the interior and assign them additional indices.
Therefore, let us take the next important step and take into account the interior.
To streamline the discussion in this section, we consider the following approach: we first address the continuum case, and subsequently generalize the results to discrete systems.

We consider a model similar to the one in the previous section, where spins take values on the 2-sphere ({$\mathbf{m} \in \mathbb{S}^2$}). Along the simple closed loop, however, the spins are restricted to the subspace ${\mathbb{S}^2 \setminus \{P_1, P_2\}}$, as illustrated in Fig.~\ref{fig_latt_and_map}(d). 
For definiteness, but without loss of generality, we assume -- similar to Sec.~\ref{sec_product} -- that the $z$-axis represents the hard axis for the spins, and thus the removed points are ${P_{1,2} = \pm \hat{\mathbf{e}}_z}$. 
Extending this framework to arbitrary positions of the points (${P_1 \neq P_2}$) on the spin sphere, including cases where they are close to each other, is straightforward.

\begin{figure}[!ht]
	\includegraphics[width=8.4cm]{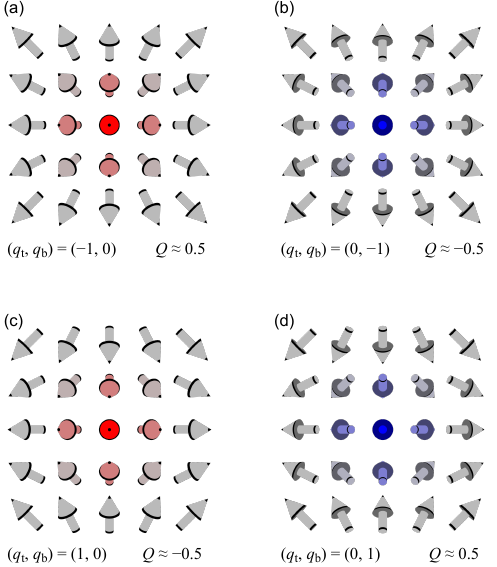}
\caption{\label{fig_merons}
Sketches of elementary merons viewed from the~$+z$~direction. 
The topological charge for each state is represented by an ordered pair of integers~$(q_\text{t}, q_\text{b})$. 
The definitions of~$q_\text{t}$ and~$q_\text{b}$ are done in the text Eq.~\eqref{qtqb} and~$Q(\mathbf{m})$ is defined in Eq.~\eqref{Q}.
}
\end{figure}

We use the notation where $\Omega$ represents the spatial domain, which is homeomorphic to a disk or a square ($\Omega \cong I^2$), and the boundary of this domain, $\partial\Omega$, is a loop homeomorphic to a circle ($\partial\Omega \cong \mathbb{S}^1$).
Then, the following continuous map takes place: 
\begin{equation}
f: (\Omega, \partial\Omega) \rightarrow (\mathbb{S}^2, \mathbb{S}^2\backslash \{ P_1, P_2\}).
\label{map_f}
\end{equation}
For the corresponding homotopy classes we found the group (for details see~\ref{app_rel_hom}): 
\begin{equation}
\pi_2(\mathbb{S}^2, \mathbb{S}^2\backslash \{ P_1, P_2\}) = \mathbb{Z}\times\mathbb{Z}.  
\label{pi2_main}
\end{equation}
An important property of the group isomorphism in Eq.~\eqref{pi2_main} is that the sum operation in the group~$\mathbb{Z}\times\mathbb{Z}$ corresponds to the standard concatenation in relative homotopy groups~\cite{Hatcher, DubrovinFomenkoNovikov_2}.
Since ${\mathbb{Z}\times\mathbb{Z}\ncong\mathbb{Z}}$ in the category of groups, the distinct homotopy types in the group Eq.~\eqref{pi2_main} are described by an ordered pair of integers rather than a single integer. 
This naturally raises the question: What invariants correspond to each of these integers, and how can one compute them? 
Although it is straightforward to take the winding number~$w(\mathbf{m})$ as the first integer and compute it using Eq.~\eqref{winding}, the origin of the second integer is less obvious. 
Thus, in the literature, spin textures similar to those shown in Fig.~\ref{fig_latt_and_map}(c) are frequently analyzed by computing the Kronecker integral~\cite{Kronecker_1869, Flanders_DifferentialForms, Mapping_degree_theory}:
\begin{equation}
Q(\mathbf{m}) = \frac{1}{4\pi}\int\limits_\Omega dr_1 dr_2\ 
\mathbf{m}\cdot[\partial_{r_1}{\mathbf{m}} \times\partial_{r_2}{\mathbf{m}}],
\label{Q}
\end{equation}
see for instance Refs.~\cite{Senthil_2004, PhysRevB.75.012408, PhysRevLett.99.117202, PhysRevB.91.224407, Augustin2021, Strungaru2022}.
However, this expression naturally gives non-integer values because it is a topological charge only for $\pi_2(\mathbb{S}^2)$ classes. 
To illustrate this, in Fig.~\ref{fig_merons} we show the configurations of elementary merons supplemented with the corresponding representative values of ${Q(\mathbf{m})\approx \pm 0.5}$ calculated for a specially selected domain~$\Omega$ and a local coordinate system $(r_1,r_2,r_3)$ with polarity $\hat{\mathbf{e}}_{r_3}=\hat{\mathbf{e}}_{z}$. 
Although Equation~\eqref{Q} will prove useful later, at this point it is essential to note that, in general, for an arbitrary spin configuration (a black-box texture) where~$Q(\mathbf{m})$, or a pair of~$w(\mathbf{m})$ and~$Q(\mathbf{m})$, are known, it is not possible to determine the homotopy type of the map Eq.~\eqref{map_f}.
In other words, $Q(\mathbf{m})$ does not serve as a topological invariant for the group in Eq.~\eqref{pi2_main}. 
This statement is proven in~\ref{app_coex} using the counterexample method.

\begin{figure}[!h]
	\includegraphics[width=7.8cm]{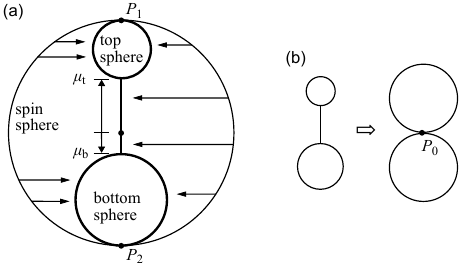}
\caption{\label{fig_wedge_sum_spheres}
Sketch illustrating the continuous transformation of a sphere into two spheres connected at a point. 
(a) Projection of the sphere onto the inscribed dumbbell. 
(b) Contraction of the handle to point, and the magnification of spheres.
}
\end{figure}

To make the computation of topological indices for the group in Eq.~\eqref{pi2_main} straightforward, we introduce a specially designed auxiliary map in spin space: 
\begin{equation}
p: (\mathbb{S}^2,\mathbb{S}^2\backslash\{ P_1, P_2 \}) \rightarrow (\mathbb{S}^2\vee\mathbb{S}^2, P_0),
\label{map_p}
\end{equation}
where the image is the wedge sum of spheres and $P_0$ is its common point, see Fig.~\ref{fig_wedge_sum_spheres}. 
The surjective map~$p$ is the essence of the two-step procedure. 
In the first step, the spin sphere is projected onto the inscribed ``dumbbell'', see Fig.~\ref{fig_wedge_sum_spheres}(a). 
To perform the projection, we drop a perpendicular line from a point on the spin sphere to the $P_1 P_2$-axis and find the intersection of this perpendicular line with either the top sphere, the bottom sphere, or the handle of the dumbbell, depending on the location of the point. 
The sizes of the \textit{top} and \textit{bottom} spheres depend on the shape parameters $\mu_\text{t}$ and $\mu_\text{b}$, respectively. 
The following relation must hold: $-1\leq \mu_\text{b} \leq \mu_\text{t} \leq 1$.
In the second step, the ``handle'' contracts to point~$P_0$, while inscribed top and bottom spheres expand into unit spheres, see Fig.~\ref{fig_wedge_sum_spheres}(b). 
The resulting construction is identical to gluing two spheres together when considering the sphere as a so-called co-H-space~\cite{ModernClassicalHomotopyTheory}, but it plays a different role here.
The continuous connection between elements of pairs in the Eq.~\eqref{map_p} is the essence of the homotopy class with the following representative. 
For parameter ${\tau\in [0,1]}$, the shape parameters evolve as: ${\mu_\text{t} = \tau}$, ${\mu_\text{b} = -\tau}$.

A key aspect of the map~$p$ in Eq.~\eqref{map_p} is that the homotopy group of its image has been thoroughly studied~\cite{Hilton1955, BottTu1982}. 
The later is because the homotopy groups of bouquets of spheres are essential for Whitehead product and homotopy constructions. 
Specifically, the second homotopy group of a bouquet of two spheres is 
\begin{equation}
\pi_2(\mathbb{S}^2\vee\mathbb{S}^2, P_0) = \mathbb{Z}\times\mathbb{Z}.
\label{pi2_S2vS2}
\end{equation}
In addition, the homomorphism of the relative homotopy groups Eq.~\eqref{pi2_main} and Eq.~\eqref{pi2_S2vS2} induced by the map in Eq.~\eqref{map_p} turns out to be an automorphism -- isomorphism to itself (for details see~\ref{app_auto}). 
Consequently, we conclude that the pair of integers $(q_\text{t}, q_\text{b})$ -- referred to as the \textit{top} and \textit{bottom} charges in Eq.~\eqref{pi2_S2vS2} -- serves as the complete invariant for the objective map in Eq.~\eqref{map_f}.
It is straightforward to calculate these integers using the degree formula~\eqref{Q} for the unit vectors~$\mathbf{m}_\text{t}$ and~$\mathbf{m}_\text{b}$ emerging from the unit vector~$\mathbf{m}$ through the map~$p$. 
From the structure of map~$p$ we obtain explicit formulas: 
\begin{equation}
\mathbf{m}_\text{t} = 
\begin{pmatrix}
2{\gamma_\text{t}} m_x  \\
2{\gamma_\text{t}} m_y  \\
- 1 + 2{\gamma_\text{t}}^2(1 + m_z)(1 - \mu_\text{t}) 
\end{pmatrix}, \label{m_t}
\end{equation}
where $\gamma_\text{t} = 
\begin{cases}
\sqrt{\dfrac{m_z - \mu_\text{t}}{(1 + m_z)(1 - \mu_\text{t})^2}}\quad\text{if}\quad \mu_\text{t}<m_z, \\
0\quad\text{otherwise}, 
\end{cases}$
and
\begin{equation}
\mathbf{m}_\text{b}  = 
\begin{pmatrix}
2{\gamma_\text{b}} m_x  \\
2{\gamma_\text{b}} m_y  \\
1 - 2{\gamma_\text{b}}^2(1 - m_z)(1 + \mu_\text{b}) 
\end{pmatrix}, \label{m_b}
\end{equation}
where
$\gamma_\text{b} = 
\begin{cases}
\sqrt{\dfrac{\mu_\text{b} - m_z}{(1 - m_z)(1 + \mu_\text{b})^2}}\quad\text{if}\quad m_z<\mu_\text{b}, \\
0\quad\text{otherwise}. 
\end{cases} \nonumber
$
The shape parameters $\mu_\text{t}$ and $\mu_\text{b}$ must be a continuous function of the coordinate in the $\Omega$ domain. One can choose monotonic functions such that $\mu_\text{t} = \mu_\text{b} = 0$ in the center of~$\Omega$ and $\mu_\text{t}=1$, $\mu_\text{b}=-1$ everywhere at the boundary~$\partial\Omega$. 
The topological charges are computed as 
\begin{equation}
q_\text{t} = Q(\mathbf{m}_\text{t}),\quad \text{and}\quad q_\text{b} = Q(\mathbf{m}_\text{b}), 
\label{qtqb}
\end{equation}
where the polarities are set to~$\hat{\mathbf{e}}_{r_3}=-\hat{\mathbf{e}}_{z}$ for $\mathbf{m}_\text{t}$ and $\hat{\mathbf{e}}_{r_3}=\hat{\mathbf{e}}_{z}$ for $\mathbf{m}_\text{b}$. 
Such a choice of polarity for the coordinate frames ensures alignment with the unit vectors corresponding to the basepoint and eliminates ambiguities discussed in Refs.~\cite{Rybakov_thesis, Zheng2023}. 
Thus, when the global coordinate system is flipped upside-down, the \textit{top} and \textit{bottom} indices naturally interchange. 
On the other hand, the topological invariant Eq.~\eqref{winding} does not depend on $z$-polarity and, accordingly, the following fundamental relation holds: 
\begin{align}
w = -(q_\text{t} + q_\text{b}).
\label{w_from_q}
\end{align}
Due to the automorphisms of the group $\mathbb{Z}\times\mathbb{Z}$ and Eq.~\eqref{w_from_q}, it is possible to derive alternative representations of the ordered pair of integers where one of them is~$w$ (for details, see~\ref{app_auto_w}). 
However, such representations result in a less symmetric framework compared to the use of $(q_\text{t}, q_\text{b})$, which we follow in this work. 
Calculation of topological charge, $(q_\text{t}, q_\text{b})$, for the configurations in Fig.~\ref{fig_merons} proves that for the group Eq.~\eqref{pi2_main}, this quartet represents two generators and their inverses. 
This is in line with earlier proposals to consider these four merons (up to trivial rotations) as a basic set of configurations~\cite{VanWaeyenberge2006, Yu2018, Gao2019}.

It also seems essential to discuss a transition between a system with two hard directions and a system with a single hard direction. 
Such a transition, for instance, may occur when the applied field starts dominating the  anisotropy~\cite{Barton-Singer2020, Ohara2022}. 
It can be associated with a transformation from $\pi_2(\mathbb{S}^2,\mathbb{S}^2\backslash\{ P_1, P_2 \})$ merons to $\pi_2(\mathbb{S}^2)$ skyrmions (see~\ref{app_rel_hom_1} and~\ref{app_rel_hom_0}). 
Since a two-punctured sphere is a subspace of a one-punctured sphere, the natural embeddings 
\begin{equation}
\begin{aligned}
&\mathbb{S}^2 \backslash \{P_1, P_2\}  \hookrightarrow  \mathbb{S}^2 \backslash \{P_2\}, \\
&\mathbb{S}^2 \backslash \{P_1, P_2\}  \hookrightarrow  \mathbb{S}^2 \backslash \{P_1\}
\end{aligned}
\end{equation}
induce group homomorphisms~$\mathbb{Z}\times\mathbb{Z}\rightarrow\mathbb{Z}$. 
These homomorphisms are the following trivial operations: 
\begin{equation}
\begin{aligned}
&(q_\text{t},q_\text{b}) \mapsto q_\text{t}, \\
&(q_\text{t},q_\text{b}) \mapsto q_\text{b}.
\end{aligned}
\label{forget}
\end{equation}
The discarding of one component of the topological invariant can be attributed to the critical transition of the system.
From the above, one can conclude that without knowledge of the hard directions in the physical system (without knowing the Hamiltonian), the calculation of the indices $q_\text{t}$ and $q_\text{b}$ becomes less meaningful, as these numbers are not guaranteed to be invariants.
From~\eqref{forget} it also follows that the number of skyrmions/antiskyrmions within the meron configuration is equal to~$\min(|q_\text{t}|,|q_\text{b}|)$.

The above approach can also be extended to the topological classification of spin textures in discrete systems. 
However, deriving a discrete version is a bit of a hassle. 
First of all, it is important to note that \textit{Axiom}~II is necessary, but not sufficient in this context.
To establish sufficient conditions, one can employ a standard approach to construct correspondences between polyhedra and  manifolds~\cite{Hatcher, Mapping_degree_theory}.  
The sufficient condition can be stated as follows: for vectors~$\mathbf{m}_{\text{t,b}}$ on the triangulation of~$\Omega$, no triples with undefined solid angles should occur. 
When this condition is satisfied, the topological indices~$q_{\text{t,b}}$ can be computed using the Berg-L\"{u}scher method~\cite{BERG1981412}.

\section{\label{sec_vector}  Secondary indices of vector fields}

In this section, we demonstrate how our unified approach can contribute to the analysis of vector fields from a general geometric perspective. 
Specifically, we consider a three-component vector field~${\mathbf{V} \equiv (V_x, V_y, V_z) \in \mathbb{R}^3}$.

A standard approach to analyzing vector fields involves the classification of singular points~\cite{Fulton_AlgebraicTopology, Mapping_degree_theory}.
Such singularities include points of zero field values. 
In the first step, one can locate and classify all such usual singularities.
As a result, in the remaining space (free from singularities), the vector field satisfies~${\mathbf{V} \in \mathbb{R}^3 \setminus \{0\}}$. 
We suggest that the analysis of such a field can be further extended by following the approach described below.
In any singularity-free simple domain~$\Omega$, it becomes possible to define indices that characterize points where the vector field aligns with \textit{specific directions}. 
The classification of such points provides additional (secondary level) information about the field.
To distinguish these indices from the indices used to classify singular points, we use the term \textit{secondary indices}.  
As before, the specific directions are denoted by points on a sphere: ${P_1, \dots, P_{r+1}}$ (${r \geq 0}$). 
As a result, all the conditions for classification via homotopy classes are satisfied because 
\begin{equation}
\mathbb{S}^2\backslash\{P_1,  ..., P_{r+1}\}\!\times\!\mathbb{R} 
\ \subset \ 
\mathbb{S}^2\!\times\!\mathbb{R}
\ \cong\ 
\mathbb{R}^3 \backslash\{0\},
\end{equation}
see Eqs.~\eqref{A_product}-\eqref{B_product} in~\ref{app_rel_hom_g}. 
Accordingly, we define the secondary index as an element of the homotopy group~{$\mathbb{Z}\times F_r$}. 
This index can take the form of an integer when ${r=0}$, a pair of integers, when ${r=1}$, or a more complex structure for higher values of~$r$.  
This general framework naturally includes the scenario described in the previous section. 
However, it is especially interesting to consider additional cases where the vector field is not constrained to the 2-sphere.
Representative examples of such systems include magnets~\cite{PhysRevB.55.3050, Ulrich_Alex_Christian_2006, Aharoni_book, Chubykalo-Fesenko2020} described by Landau-type theories and ferroelectrics~\cite{Lines_Glass_Ferroelectrics, PhysRevB.74.104104, Das2019, TopoFerroChiral}.  
Landau-type potentials penalize zero field values, enabling primary classification through usual indices. 
When there are hard directions, the classification can be extended using secondary indices, allowing for the classification of skyrmions and merons in these systems.  
Remarkably, for the practical case where one or two hard directions lie along the same axis, the formulas from the previous section can be applied directly to the field~$\mathbf{m}=\mathbf{V}/|\mathbf{V}|$.
Non-abelian cases, with three or more hard directions~\cite{Rybakov_Eriksson}, will be addressed in more detail elsewhere.


\section{\label{sec_disc} Discussion}

\begin{figure}[!h]
	\includegraphics[width=7.8cm]{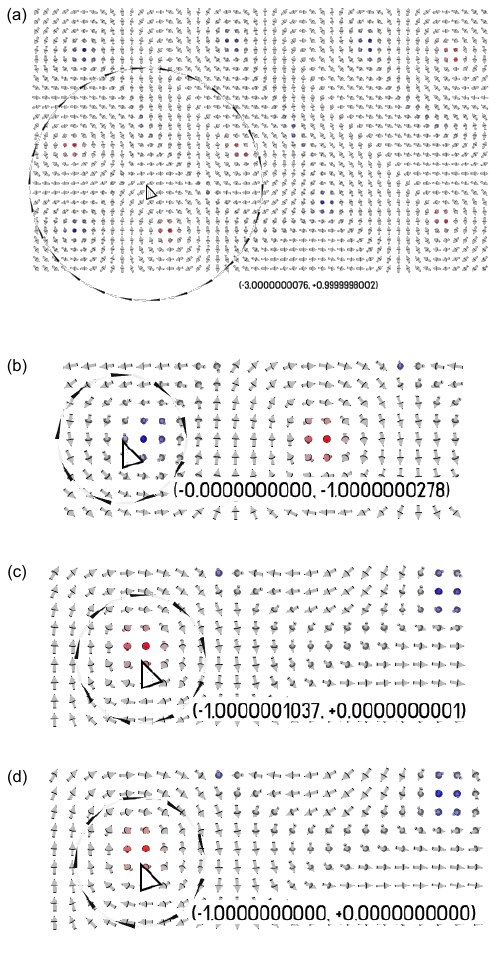}
\caption{\label{fig_scan}
Real-time topological charge calculations. 
(a)~Snapshot of the whole system, consisting of ${48\times28}$ spins.
(b)-(d)~Zoomed-in areas. 
The two numbers in each panel correspond to the values of~$q_\text{t}$ and~$q_\text{b}$ inside the circle, respectively.
}
\end{figure}

As an illustration of practical application, we applied formulas~\eqref{m_t}-\eqref{m_b} for typical, composite magnetic textures featuring merons and skyrmions~\cite{PhysRevB.91.224407, PhysRevB.101.064408}. 
We chose a setup that did not favor smoothness, intending to demonstrate robustness. 
The maximum angle between adjacent spins on the lattice reached values of about~75$^\circ$.
The profiles of parameters~$\mu_\text{t}$ and~$\mu_\text{b}$ were set as step functions, changing sharply from~${\mu_\text{t}=0}$, ${\mu_\text{b}=0}$ to~${\mu_\text{t}=1}$, ${\mu_\text{b}=-1}$ when crossing the contour boundary from the inside to the outside.
The indices~$q_{\text{t}}$ and~$q_{\text{b}}$ were calculated by the Berg-L\"{u}scher method~\cite{BERG1981412} using vectors obtained from Eqs.~\eqref{m_t}-\eqref{m_b}.
Figure~\ref{fig_scan} shows snapshots corresponding to various positions and sizes of the measuring contour (see also the Supplementary movie~\cite{suppl}).
As can be seen, the calculated indices are integers with at least six digits of precision. 
A slight discrepancy arises exclusively due to the roundoff error for 32-bit floating-point arithmetic.
To confirm that, the specialized snapshot in Fig.~\ref{fig_scan}(d) demonstrates results obtained with 64-bit floating-point arithmetic, offering a precision several orders of magnitude higher. 
This small numerical study illustrates two important aspects of our theory: 
(i) binocular understanding of the topology of vector fields rather than monocular through values of~$Q(\mathbf{m})$; 
(ii) getting rid of the need for idealized boundary conditions to identify topological invariants.

In conclusion, this work presents a unified approach for the topological classification of spin textures, grounded in simple axioms that can be objectively accepted or rejected based on physical reasoning. 
Without loss of generality, we focused on one of the most practical cases: systems where the axioms hold due to the presence of a hard spin-orientation axis.  
For this case, we derived a group consisting of topological invariants~$(q_\text{t}, q_\text{b})$ and developed a robust method for their computation using a specially designed auxiliary map. 
The result is directly applicable to ferromagnets and ferroelectrics, and can naturally be generalized to antiferromagnets~\cite{Chmiel2018, Jani2021, Amin2023, GOBEL20211}.
Notably, the presented approach extends naturally beyond spin systems, offering a versatile framework for analyzing complex topological textures and deepening our understanding of the physical properties of the systems that host them.


\setcounter{section}{0}

\renewcommand{\thesection}{\text{APPENDIX }\Alph{section}}
\renewcommand{\thesubsection}{\arabic{subsection}}
\renewcommand{\theequation}{\Alph{section}\arabic{equation}}

\section{\label{app_rel_hom} Derivation of relative homotopy group}
\setcounter{equation}{0}

In this section, we derive relative homotopy groups classifying skyrmions and vortices/merons in generalized cases, that is, for both abelian and non-abelian vortices~\cite{Rybakov_Eriksson}.

\subsection{\label{app_rel_hom_g} General case}

We consider the long exact sequence of homotopy and relative homotopy groups for the pair of topological spaces~\cite{HuHomotopyTheory, GeorgeWhitehead1979}: 
\begin{align}
\label{sequence_long}
... 
\rightarrow 
\pi_2(B) 
\rightarrow 
 \pi_2(A) & 
\xrightarrow{j_2}
 \pi_2(A, B) 
\rightarrow \\ \nonumber
& \rightarrow  \pi_1(B) 
\rightarrow 
\pi_1(A) 
\rightarrow 
...
\end{align}
We highlight one of the homomorphisms with the symbol~$j_2$ since we will pay special attention to this element of the chain complex later in the text. 
Equation~\eqref{sequence_long} is valid for any connected space~$A$ and its connected subspace ${B \subseteq A}$, whereas in our case, and in accordance with the map~\eqref{map_f}, the spaces are as follows: 
\begin{align}
A = \mathbb{S}^2, \quad 
B = \mathbb{S}^2\backslash\{P_1,  ..., P_{r+1}\}, 
\end{align}
where ${r\geq 0}$.  
We assume that~$B$ (and therefore also~$A$) is a pointed space, whereas we omit the entry of the corresponding basepoint in the formulas due to standard convention in the pointed category~\cite{HuHomotopyTheory, ModernClassicalHomotopyTheory}.
The derivation of the first and second homotopy groups involved in the sequence~\eqref{sequence_long} is straightforward~\cite{HuHomotopyTheory, Hatcher, ClimenhagaKatok}: 
\begin{align}
\pi_2(B)=0, \ 
\pi_2(A)=\mathbb{Z}, \ 
\pi_1(B)=F_r, \ 
\pi_1(A)=0,
\label{pi_2_pi_1}
\end{align}
where $F_r$ -- free group of rank~$r$. 
Substituting the results into long exact sequence~\eqref{sequence_long} we obtain a short exact sequence:
\begin{equation}
0 
\rightarrow 
\mathbb{Z} 
\xrightarrow{j_2}
\pi_2(A, B) 
\rightarrow 
F_r
\rightarrow 
0 
\label{sequence_short}. 
\end{equation}
The exactness of the short sequence~\eqref{sequence_short} is a necessary condition for the homotopy group~$\pi_2(A, B)$. 
To further solve the problem, we first find all possible solutions for group~$G$ placed at the center of sequence~\eqref{sequence_short} and for arbitrary homomorphisms preserving the exactness of the sequence: 
\begin{equation}
0 
\rightarrow 
\mathbb{Z} 
\rightarrow 
G 
\rightarrow 
F_r
\rightarrow 
0 
\label{sequence_short_G}. 
\end{equation}
From an algebraic point of view, the exactness of sequence~\eqref{sequence_short_G} means that~$G$ is an extension of~$F_r$ (image) by~$\mathbb{Z}$ (kernel)~\cite{CohomologyOfGroups}. 
Since the image is a free group, this extension splits~\cite{IntroductiontoHomological}, and only trivial extensions are possible: 
\begin{equation}
G = \mathbb{Z} \rtimes_\varphi F_r, 
\label{group_G} 
\end{equation}
where the semidirect product is defined by homomorphism 
\begin{equation}
\varphi: F_r \rightarrow \text{Aut}(\mathbb{Z}). 
\label{varphi}
\end{equation}
Semidirect products defined by a nontrivial homomorphism sometimes occur in homotopy groups. 
Some examples are the fundamental group of the Klein bottle~\cite{Kosniowski}, $\pi_1(\mathbb{RP}^2\texttt{\#}\mathbb{RP}^2)=\mathbb{Z} \rtimes\mathbb{Z}$, and the Abe homotopy constructions~\cite{Abe1940, KOBAYASHI2012577, PhysRevB.101.195130}.
We have to determine whether~$\varphi$ is trivial for the objective group~$\pi_2(A, B)$.

Only several possible homomorphisms from the Eq.~\eqref{varphi} exist.  
The automorphism group of integers, 
\begin{equation}
\text{Aut}(\mathbb{Z})\cong\mathbb{Z}_2, 
\label{aut_Z}
\end{equation}
consists of two elements corresponding to multiplications by either ``$+1$'' (trivial) or ``$-1$''. 
Therefore, for~$\varphi$ in Eq.~\eqref{group_G} there are~$2^r$ possible options so that each of the generators (and its inverse) in~$F_r$ acts on the elements of~$\mathbb{Z}$ either trivially or by changing the sign. 
Group~$\pi_2(A, B)$ must be one of the groups~$G$ from Eq.~\eqref{group_G}, and to identify it unambiguously, next we use an additional property of relative homotopy groups following from its geometric construction~\cite{Hilton1953, GeorgeWhitehead1979}: in the sequence~\eqref{sequence_long} the image of map~
$\pi_2(A)\xrightarrow{j_2}\pi_2(A, B)$ 
belongs to the center of~$\pi_2(A, B)$, i.e. 
\begin{equation}
\text{im}\,j_2 \subseteq Z(\pi_2(A, B)). 
\end{equation}
Also, from Eqs.~\eqref{sequence_short}-\eqref{group_G} it follows that 
${\text{im}\,j_2 = \mathbb{Z}\times\{0\}}$. 
However, the center of $G$ includes ${\mathbb{Z}\times\{0\}}$ only in one case -- when the homomorphism~$\varphi$ is trivial.  
Indeed, because if~$\varphi$ is non-trivial, then $Z(G)=\{0\}\times 2\mathbb{Z}$ for $r=1$ and $Z(G)=\{0\}\times \{0\}$ for ${r \geq 2}$. 
Thus, all non-trivial homomorphisms~$\varphi$ are excluded, and we obtain the final solution which is the (usual) direct product: 
\begin{equation}
\pi_2(A, B) = \mathbb{Z} \times F_r. 
\label{pi2_general}
\end{equation}

It is also useful to somehow generalize the spaces~$A$ and~$B$ for which Eq.~\eqref{pi2_general} is valid. 
Considering that the derivation of Eq.~\eqref{pi2_general} begins with the Eqs.~\eqref{pi_2_pi_1}, we can equally assume: 
\begin{equation}
A \cong \mathbb{S}^2\times X_A, 
\label{A_product}
\end{equation}
and
\begin{align}
& B \cong \mathbb{S}^2\backslash\{P_1,  ..., P_{r+1}\}\times X_B 
\quad\text{or} \label{B_product}\\ 
& B \cong \bigvee_{r}\mathbb{S}^1\times X_B, 
\end{align}
where~$X_{A,B}$ are any 2-simply connected spaces (i.e.~$\pi_{k}(X_i)=0$ for ${k\leq 2}$), symbol~$\bigvee$ stands for the bouquet~\cite{Hatcher, ClimenhagaKatok}, and ${B \subset A}$.

The relative homotopy group~\eqref{pi2_general} is abelian (commutative) if and only if~$r$ is either~$0$ or~$1$. 


\subsection{\label{app_rel_hom_0} Abelian group for $r=0$ and its automorphisms}

For ${r=0}$, the group Eq.~\eqref{pi2_general} is the integers: 
\begin{equation}
\pi_2(\mathbb{S}^2,  \mathbb{S}^2\backslash\{ P_1\}) = 
\mathbb{Z} \times \{ 0 \} = \mathbb{Z}. 
\label{pi2_S2}
\end{equation}
Since the space~$\mathbb{S}^2\backslash\{ P_1\}$ is contractible, it can be replaced by a point in the Eq.~\eqref{pi2_S2} giving the usual second homotopy group of the sphere~\cite{Hilton1953, HuHomotopyTheory}: ${\pi_2(\mathbb{S}^2, P_0) = \pi_2(\mathbb{S}^2) = \mathbb{Z}}$.

The automorphism group of integers is given in the Eq.~\eqref{aut_Z}. 
The two-element nature of this group is the essence of the ambiguity in the choice of sign. 
Thus, one can adopt one of two possible conventions for the sign of a skyrmion.
For example, the authors of Refs.~\cite{KOSEVICH1990117, PAPANICOLAOU1991425, PhysRevResearch.5.043199} assumed that N\'{e}el or Bloch skyrmions have a positive charge.
Alternatively, in Refs.~\cite{Melcher_2014, Hoffmann2017, PhysRevB.95.094423, Barton-Singer2020}, the above sign is assumed to be negative, and we also follow this convention.

\subsection{\label{app_rel_hom_1} Abelian group for $r=1$ and its automorphisms}

For ${r=1}$, the group Eq.~\eqref{pi2_general} is free abelian of rank~2:
\begin{equation}
\pi_2(\mathbb{S}^2, \mathbb{S}^2\backslash\{ P_1, P_2 \}) = 
\mathbb{Z} \times F_1 = \mathbb{Z} \times \mathbb{Z}. 
\label{pi2_S2_S1}
\end{equation}
It is worth noting that since the circle~$\mathbb{S}^1$ is a deformation retract for the space $\mathbb{S}^2\backslash\{ P_1, P_2 \}$, there is an isomorphism that immediately gives from Eq.~\eqref{pi2_S2_S1} the following constrained case: ${\pi_2(\mathbb{S}^2, \mathbb{S}^1) = \mathbb{Z}\times\mathbb{Z}}$. 
In several works, the construction ${\pi_2(\mathbb{S}^2, \mathbb{S}^1)}$ has been explicitly mentioned before as being related to vortices/merons, but has not been derived or applied~\cite{PhysRevB.85.144433, Nagase2021}. 
Also, the derivation of the group~${\pi_2(\mathbb{S}^2, \mathbb{S}^1)}$ in the context of other applications may be found in Ref.~\cite{RicardoKennedyThesis2014}.

The automorphism group for Eq.~\eqref{pi2_S2_S1} is the general linear group of degree~2 over integers:
\begin{equation}
\text{Aut}(\mathbb{Z}\times\mathbb{Z})\cong\text{GL}_2(\mathbb{Z}).
\label{aut_ZxZ}
\end{equation}
This group is infinite and is much more complex than the group~\eqref{aut_Z}. 
Accordingly, establishing a convention for selecting the pair of indices is particularly important.

\section{\label{app_auto} Automorphism induced by the auxiliary map}
\setcounter{equation}{0}

In this section, we show that the auxiliary map~\eqref{map_p} induces an automorphism of the corresponding homotopy group.

The continuous map of any pair $(A,B)$, ${B\subseteq A}$, into a pair $(C,D)$, ${D\subseteq C}$, naturally induces a homomorphism of the corresponding relative homotopy groups~\cite{Hilton1953, DubrovinFomenkoNovikov_2}: $\pi_2(A, B)\rightarrow\pi_2(C, D)$. 
In particular, if two images in $\pi_2(C, D)$ have different homotopy types, then their preimages in $\pi_2(A, B)$ must have different homotopy types.
It can be illustrated with the following diagram, which becomes commutative for the trivial composition rule~${h = p \circ f}$: 
\begin{equation}
\begin{tikzcd}[column sep=large]
{}  & (A, B) \arrow{d}{p} \\
(\Omega,\partial\Omega) \arrow[dashed]{r}{h}  \arrow{ru}{f} & (C, D).
\end{tikzcd}
\label{diag}
\end{equation}

Summarizing the above and taking into account the Eqs.~\eqref{pi2_main} and~\eqref{pi2_S2vS2}, we obtain the following diagram: 
\begin{equation}
\begin{tikzcd}
\pi_2(\mathbb{S}^2, \mathbb{S}^2\backslash\{ P_1, P_2 \}) \arrow[r] \arrow[equal, d]
&  \pi_2(\mathbb{S}^2 \vee \mathbb{S}^2, P_0)  \arrow[equal, d ] \\
\mathbb{Z}\times\mathbb{Z} \arrow[r] 
& \mathbb{Z}\times\mathbb{Z},  
\end{tikzcd} 
\label{homom_diag}
\end{equation}
induced by the map~\eqref{map_p}. 
Taking into account that all possible self-homomorphisms of~${\mathbb{Z}\times\mathbb{Z}}$ form the following semigroup: 
\begin{equation}
\text{Hom}(\mathbb{Z}\times\mathbb{Z},\mathbb{Z}\times\mathbb{Z}) \equiv \text{End}(\mathbb{Z}\times\mathbb{Z}) \cong \text{M}_2(\mathbb{Z}),
\label{homom}
\end{equation}
we find that the integer invariants of both groups in the diagram~\eqref{homom_diag} are necessarily linearly related by~${2\!\times\!2}$~matrix with integer coefficients, i.e. 
\begin{equation}
M \cdot 
\begin{pmatrix}
q_\text{t}^{\prime}\\
q_\text{b}^{\prime}
\end{pmatrix} = 
\begin{pmatrix}
q_\text{t}\\
q_\text{b}
\end{pmatrix}, \quad \text{where} \quad M \in \text{M}_2(\mathbb{Z}), 
\label{M2}
\end{equation}
and the symbols $^\prime$ denote belonging to the left side of the diagram~\eqref{homom_diag}. 
It is now straightforward to determine the type of homomorphism even more specifically than Eq.~\eqref{M2} gives, by examining the mapping on some given configurations. 
In other words, it makes sense to check homotopy types for~$h$ in the diagram~\eqref{diag} while trying different~$f$, and from this to extract more information about the homomorphism. 
Calculations of topological charges $(q_\text{t},q_\text{b})$ for the merons illustrated in Figs.~\ref{fig_merons}(c) and~(d) give~$(1,0)$ and~$(0,1)$, respectively. 
Hence, according to Eq.~\eqref{M2}, there exist four integers, $i_{1,2,3,4}$, such that 
\begin{equation}
M \cdot 
\begin{pmatrix}
i_1\\
i_2
\end{pmatrix} = 
\begin{pmatrix}
1\\
0
\end{pmatrix}, \quad \text{and} \quad 
M \cdot 
\begin{pmatrix}
i_3\\
i_4
\end{pmatrix} = 
\begin{pmatrix}
0\\
1
\end{pmatrix}. 
\label{i1i2i3i4}
\end{equation}
These two resulting equations can be represented by one matrix equation: 
\begin{equation}
M \cdot 
\begin{pmatrix}
i_1 & i_3\\
i_2 & i_4
\end{pmatrix} = 
\begin{pmatrix}
1 & 0\\
0 & 1
\end{pmatrix}. 
\end{equation}
In this way, we found that for~$M$, there is an inverse matrix with integer coefficients. 
Combining this fact with the fact that the coefficients of~$M$ are integers, we get: 
\begin{equation}
M\in\text{GL}_2(\mathbb{Z}).
\label{M_inGL2}
\end{equation}

As a result we obtain that~$M$ is the element of automorphisms~\eqref{aut_ZxZ} and therefore the homomorphism~\eqref{homom_diag} belongs to the automorphisms.


\section{\label{app_auto_w} Automorphism aimed at isolating the winding number}
\setcounter{equation}{0}

In this section, we find all variants of automorphisms from Eq.~\eqref{aut_ZxZ} leading to the isolation of index~$w$ satisfying Eq.~\eqref{w_from_q}.
Using the Eq.~\eqref{w_from_q} we get the following Diophantine problem:
\begin{equation}
K \cdot 
\begin{pmatrix}
q_\text{t}\\
q_\text{b}
\end{pmatrix} = 
\begin{pmatrix}
-q_\text{t} - q_\text{b}\\
v
\end{pmatrix}, \quad K\in\text{GL}_2(\mathbb{Z}).
\label{Dioph}
\end{equation} 
All solutions of Eq.~\eqref{Dioph} can be divided into two sets, and these are either 
\begin{equation}
v = k\,w + q_\text{t}, \quad \text{or} \quad v = k\,w + q_\text{b}, 
\end{equation}
for all ${k \in \mathbb{Z}}$.
Thus, this eventually leads to an asymmetry in the sense that one has to give preference to either equation Eq.~\eqref{m_t} or Eq.~\eqref{m_b}.


\section{\label{app_coex} Useful counterexample}
\setcounter{equation}{0}

In this section, we show that homotopic configurations in group~\eqref{pi2_main} may have an arbitrarily large difference in the values of~$Q(\mathbf{m})$.

Our approach here is in the spirit of previous works~\cite{PhysRevB.91.224407, PhysRevB.101.064408, PhysRevB.110.094442}, demonstrating continuous distortions of merons and skyrmions in favor of the northern or southern spin hemisphere. 
We consider two configurations:
\begin{align}
\mathbf{M}_\pm = 
\begin{pmatrix}
\sqrt{1-g_\pm(\rho)^2} \cos(k\,\phi + \phi_0)  \\
\sqrt{1-g_\pm(\rho)^2} \sin(k\,\phi + \phi_0)  \\
g_\pm(\rho)
\end{pmatrix},
\label{m_pm}
\end{align}
where $k$ is any integer, $\phi_0$ is arbitrary parameter, and $(\rho,\phi)$ -- polar coordinates centered in circular domain~$\Omega$ with radius~$\rho_0$. 
Let the profile functions be monotone and satisfy the following conditions: 
${g_{\pm}(0) = 1}$ and ${g_{\pm}(\rho_0)=\pm m_0}$, where ${0<m_0<1}$. 
Using Eq.~\eqref{Q}, we obtain the difference 
\begin{align}
Q(\mathbf{M}_-) - Q(\mathbf{M}_+) = k\,m_0,  
\end{align}
which can be equal to any real number. 
Whereas the topological charge for both configurations Eq.~\eqref{m_pm} is identical,~${(q_\text{t},q_\text{b}) = (-k,0)}$. 
It is worth noting that both configurations Eq.~\eqref{m_pm} are suitable ansatz for merons in the system with an easy-cone spin-orientation anisotropy~\cite{Hubert, Skomski}.

\begin{acknowledgments}
We thank Anna Delin, Patrik Thunstr\"{o}m, and Egor Babaev for discussing the results. 
F.N.R. and O.E. acknowledge support from the Swedish Research Council (Grant No. 2023-04899).
O.E. acknowledges support from eSSENCE, STandUPP and the European Research Council (FASTCORR project 854843) as well as WISE -- the Wallenberg Initiative Materials Science, funded by the Knut and Alice Wallenberg Foundation. 
O.E. also acknowledges support from the Knut and Alice Wallenberg Foundation from the projects KAW 2022.0108 and KAW 2022.0252.
N.S.K. acknowledges support from the European Research Council under the European Union's Horizon 2020 Research and Innovation Programme (Grant No.~856538 -- project ``3D MAGiC'').
\end{acknowledgments}

\bibliography{bibliography.bib}

\begin{thebibliography}{135}%
\makeatletter
\providecommand \@ifxundefined [1]{%
 \@ifx{#1\undefined}
}%
\providecommand \@ifnum [1]{%
 \ifnum #1\expandafter \@firstoftwo
 \else \expandafter \@secondoftwo
 \fi
}%
\providecommand \@ifx [1]{%
 \ifx #1\expandafter \@firstoftwo
 \else \expandafter \@secondoftwo
 \fi
}%
\providecommand \natexlab [1]{#1}%
\providecommand \enquote  [1]{``#1''}%
\providecommand \bibnamefont  [1]{#1}%
\providecommand \bibfnamefont [1]{#1}%
\providecommand \citenamefont [1]{#1}%
\providecommand \href@noop [0]{\@secondoftwo}%
\providecommand \href [0]{\begingroup \@sanitize@url \@href}%
\providecommand \@href[1]{\@@startlink{#1}\@@href}%
\providecommand \@@href[1]{\endgroup#1\@@endlink}%
\providecommand \@sanitize@url [0]{\catcode `\\12\catcode `\$12\catcode `\&12\catcode `\#12\catcode `\^12\catcode `\_12\catcode `\%12\relax}%
\providecommand \@@startlink[1]{}%
\providecommand \@@endlink[0]{}%
\providecommand \url  [0]{\begingroup\@sanitize@url \@url }%
\providecommand \@url [1]{\endgroup\@href {#1}{\urlprefix }}%
\providecommand \urlprefix  [0]{URL }%
\providecommand \Eprint [0]{\href }%
\providecommand \doibase [0]{https://doi.org/}%
\providecommand \selectlanguage [0]{\@gobble}%
\providecommand \bibinfo  [0]{\@secondoftwo}%
\providecommand \bibfield  [0]{\@secondoftwo}%
\providecommand \translation [1]{[#1]}%
\providecommand \BibitemOpen [0]{}%
\providecommand \bibitemStop [0]{}%
\providecommand \bibitemNoStop [0]{.\EOS\space}%
\providecommand \EOS [0]{\spacefactor3000\relax}%
\providecommand \BibitemShut  [1]{\csname bibitem#1\endcsname}%
\let\auto@bib@innerbib\@empty
\bibitem [{\citenamefont {Malozemoff}\ and\ \citenamefont {Slonczewski}(1979)}]{MalozemoffSlonczewski_1979}%
  \BibitemOpen
  \bibfield  {author} {\bibinfo {author} {\bibfnamefont {A.~P.}\ \bibnamefont {Malozemoff}}\ and\ \bibinfo {author} {\bibfnamefont {J.~C.}\ \bibnamefont {Slonczewski}},\ }\href@noop {} {\emph {\bibinfo {title} {Magnetic Domain Walls in Bubble Materials}}}\ (\bibinfo  {publisher} {Academic Press, New York},\ \bibinfo {year} {1979})\BibitemShut {NoStop}%
\bibitem [{\citenamefont {Hubert}\ and\ \citenamefont {Schäfer}(1998)}]{Hubert}%
  \BibitemOpen
  \bibfield  {author} {\bibinfo {author} {\bibfnamefont {A.}~\bibnamefont {Hubert}}\ and\ \bibinfo {author} {\bibfnamefont {R.}~\bibnamefont {Schäfer}},\ }\href {https://doi.org/10.1007/978-3-540-85054-0} {\emph {\bibinfo {title} {Magnetic Domains: The Analysis of Magnetic Microstructures}}}\ (\bibinfo  {publisher} {Springer, Berlin, Heidelberg},\ \bibinfo {year} {1998})\BibitemShut {NoStop}%
\bibitem [{\citenamefont {DeSimone}\ \emph {et~al.}(2006)\citenamefont {DeSimone}, \citenamefont {Kohn}, \citenamefont {M{\"u}ller},\ and\ \citenamefont {Otto}}]{DeSimone_chapter4_2006}%
  \BibitemOpen
  \bibfield  {author} {\bibinfo {author} {\bibfnamefont {A.}~\bibnamefont {DeSimone}}, \bibinfo {author} {\bibfnamefont {R.~V.}\ \bibnamefont {Kohn}}, \bibinfo {author} {\bibfnamefont {S.}~\bibnamefont {M{\"u}ller}},\ and\ \bibinfo {author} {\bibfnamefont {F.}~\bibnamefont {Otto}},\ }\bibfield  {title} {\bibinfo {title} {Recent analytical developments in micromagnetics},\ }in\ \href {https://doi.org/10.1016/B978-012480874-4/50015-4} {\emph {\bibinfo {booktitle} {The Science of Hysteresis}}},\ \bibinfo {editor} {edited by\ \bibinfo {editor} {\bibfnamefont {G.}~\bibnamefont {Bertotti}}\ and\ \bibinfo {editor} {\bibfnamefont {I.~D.}\ \bibnamefont {Mayergoyz}}}\ (\bibinfo  {publisher} {Academic Press},\ \bibinfo {address} {Oxford},\ \bibinfo {year} {2006})\ pp.\ \bibinfo {pages} {269--381}\BibitemShut {NoStop}%
\bibitem [{\citenamefont {Braun}(2012)}]{BraunReview2012}%
  \BibitemOpen
  \bibfield  {author} {\bibinfo {author} {\bibfnamefont {H.-B.}\ \bibnamefont {Braun}},\ }\bibfield  {title} {\bibinfo {title} {Topological effects in nanomagnetism: from superparamagnetism to chiral quantum solitons},\ }\href {https://doi.org/10.1080/00018732.2012.663070} {\bibfield  {journal} {\bibinfo  {journal} {Advances in Physics}\ }\textbf {\bibinfo {volume} {61}},\ \bibinfo {pages} {1} (\bibinfo {year} {2012})}\BibitemShut {NoStop}%
\bibitem [{\citenamefont {Zang}\ \emph {et~al.}(2018)\citenamefont {Zang}, \citenamefont {Cros},\ and\ \citenamefont {Hoffmann}}]{TopologyMagnetism2018}%
  \BibitemOpen
  \bibinfo {editor} {\bibfnamefont {J.}~\bibnamefont {Zang}}, \bibinfo {editor} {\bibfnamefont {V.}~\bibnamefont {Cros}},\ and\ \bibinfo {editor} {\bibfnamefont {A.}~\bibnamefont {Hoffmann}},\ eds.,\ \href {https://doi.org/10.1007/978-3-319-97334-0} {\emph {\bibinfo {title} {Topology in Magnetism}}},\ Springer Series in Solid-State Sciences, Vol. 192\ (\bibinfo  {publisher} {Springer Cham},\ \bibinfo {year} {2018})\BibitemShut {NoStop}%
\bibitem [{\citenamefont {Landau}\ and\ \citenamefont {Lifshitz}(1935)}]{LL1935}%
  \BibitemOpen
  \bibfield  {author} {\bibinfo {author} {\bibfnamefont {L.~D.}\ \bibnamefont {Landau}}\ and\ \bibinfo {author} {\bibfnamefont {E.~M.}\ \bibnamefont {Lifshitz}},\ }\bibfield  {title} {\bibinfo {title} {Theory of the dispersion of magnetic permeability in ferromagnetic bodies},\ }\href@noop {} {\bibfield  {journal} {\bibinfo  {journal} {Phys. Z. Sowietunion}\ }\textbf {\bibinfo {volume} {8}},\ \bibinfo {pages} {153–169} (\bibinfo {year} {1935})}\BibitemShut {NoStop}%
\bibitem [{\citenamefont {Hertel}\ \emph {et~al.}(2005)\citenamefont {Hertel}, \citenamefont {Fruchart}, \citenamefont {Cherifi}, \citenamefont {Jubert}, \citenamefont {Heun}, \citenamefont {Locatelli},\ and\ \citenamefont {Kirschner}}]{PhysRevB.72.214409}%
  \BibitemOpen
  \bibfield  {author} {\bibinfo {author} {\bibfnamefont {R.}~\bibnamefont {Hertel}}, \bibinfo {author} {\bibfnamefont {O.}~\bibnamefont {Fruchart}}, \bibinfo {author} {\bibfnamefont {S.}~\bibnamefont {Cherifi}}, \bibinfo {author} {\bibfnamefont {P.-O.}\ \bibnamefont {Jubert}}, \bibinfo {author} {\bibfnamefont {S.}~\bibnamefont {Heun}}, \bibinfo {author} {\bibfnamefont {A.}~\bibnamefont {Locatelli}},\ and\ \bibinfo {author} {\bibfnamefont {J.}~\bibnamefont {Kirschner}},\ }\bibfield  {title} {\bibinfo {title} {Three-dimensional magnetic-flux-closure patterns in mesoscopic {Fe} islands},\ }\href {https://doi.org/10.1103/PhysRevB.72.214409} {\bibfield  {journal} {\bibinfo  {journal} {Phys. Rev. B}\ }\textbf {\bibinfo {volume} {72}},\ \bibinfo {pages} {214409} (\bibinfo {year} {2005})}\BibitemShut {NoStop}%
\bibitem [{\citenamefont {Arrott}(2011)}]{Arrott2011}%
  \BibitemOpen
  \bibfield  {author} {\bibinfo {author} {\bibfnamefont {A.~S.}\ \bibnamefont {Arrott}},\ }\bibfield  {title} {\bibinfo {title} {Dipole-dipole interactions in the computational micromagnetism of iron (1955-2010) (invited)},\ }\href {https://doi.org/10.1063/1.3561783} {\bibfield  {journal} {\bibinfo  {journal} {Journal of Applied Physics}\ }\textbf {\bibinfo {volume} {109}},\ \bibinfo {pages} {07E135} (\bibinfo {year} {2011})}\BibitemShut {NoStop}%
\bibitem [{\citenamefont {Guo}\ \emph {et~al.}(2022)\citenamefont {Guo}, \citenamefont {Henschel}, \citenamefont {Wolf}, \citenamefont {Pohl}, \citenamefont {Lubk}, \citenamefont {Blon}, \citenamefont {Neu},\ and\ \citenamefont {Leistner}}]{fromLPatterntoVortex2022}%
  \BibitemOpen
  \bibfield  {author} {\bibinfo {author} {\bibfnamefont {S.}~\bibnamefont {Guo}}, \bibinfo {author} {\bibfnamefont {M.}~\bibnamefont {Henschel}}, \bibinfo {author} {\bibfnamefont {D.}~\bibnamefont {Wolf}}, \bibinfo {author} {\bibfnamefont {D.}~\bibnamefont {Pohl}}, \bibinfo {author} {\bibfnamefont {A.}~\bibnamefont {Lubk}}, \bibinfo {author} {\bibfnamefont {T.}~\bibnamefont {Blon}}, \bibinfo {author} {\bibfnamefont {V.}~\bibnamefont {Neu}},\ and\ \bibinfo {author} {\bibfnamefont {K.}~\bibnamefont {Leistner}},\ }\bibfield  {title} {\bibinfo {title} {Size-specific magnetic configurations in electrodeposited epitaxial iron nanocuboids: From {Landau} pattern to vortex and single domain states},\ }\href {https://doi.org/10.1021/acs.nanolett.2c00607} {\bibfield  {journal} {\bibinfo  {journal} {Nano Letters}\ }\textbf {\bibinfo {volume} {22}},\ \bibinfo {pages} {4006} (\bibinfo {year} {2022})}\BibitemShut {NoStop}%
\bibitem [{\citenamefont {Shinjo}\ \emph {et~al.}(2000)\citenamefont {Shinjo}, \citenamefont {Okuno}, \citenamefont {Hassdorf}, \citenamefont {Shigeto},\ and\ \citenamefont {Ono}}]{Shinjo2000}%
  \BibitemOpen
  \bibfield  {author} {\bibinfo {author} {\bibfnamefont {T.}~\bibnamefont {Shinjo}}, \bibinfo {author} {\bibfnamefont {T.}~\bibnamefont {Okuno}}, \bibinfo {author} {\bibfnamefont {R.}~\bibnamefont {Hassdorf}}, \bibinfo {author} {\bibfnamefont {K.}~\bibnamefont {Shigeto}},\ and\ \bibinfo {author} {\bibfnamefont {T.}~\bibnamefont {Ono}},\ }\bibfield  {title} {\bibinfo {title} {Magnetic vortex core observation in circular dots of permalloy},\ }\href {https://doi.org/10.1126/science.289.5481.930} {\bibfield  {journal} {\bibinfo  {journal} {Science}\ }\textbf {\bibinfo {volume} {289}},\ \bibinfo {pages} {930} (\bibinfo {year} {2000})}\BibitemShut {NoStop}%
\bibitem [{\citenamefont {Raabe}\ \emph {et~al.}(2000)\citenamefont {Raabe}, \citenamefont {Pulwey}, \citenamefont {Sattler}, \citenamefont {Schweinböck}, \citenamefont {Zweck},\ and\ \citenamefont {Weiss}}]{Raabe2000}%
  \BibitemOpen
  \bibfield  {author} {\bibinfo {author} {\bibfnamefont {J.}~\bibnamefont {Raabe}}, \bibinfo {author} {\bibfnamefont {R.}~\bibnamefont {Pulwey}}, \bibinfo {author} {\bibfnamefont {R.}~\bibnamefont {Sattler}}, \bibinfo {author} {\bibfnamefont {T.}~\bibnamefont {Schweinböck}}, \bibinfo {author} {\bibfnamefont {J.}~\bibnamefont {Zweck}},\ and\ \bibinfo {author} {\bibfnamefont {D.}~\bibnamefont {Weiss}},\ }\bibfield  {title} {\bibinfo {title} {Magnetization pattern of ferromagnetic nanodisks},\ }\href {https://doi.org/10.1063/1.1289216} {\bibfield  {journal} {\bibinfo  {journal} {Journal of Applied Physics}\ }\textbf {\bibinfo {volume} {88}},\ \bibinfo {pages} {4437} (\bibinfo {year} {2000})}\BibitemShut {NoStop}%
\bibitem [{\citenamefont {Guslienko}\ \emph {et~al.}(2001)\citenamefont {Guslienko}, \citenamefont {Novosad}, \citenamefont {Otani}, \citenamefont {Shima},\ and\ \citenamefont {Fukamichi}}]{Guslienko_2001}%
  \BibitemOpen
  \bibfield  {author} {\bibinfo {author} {\bibfnamefont {K.~Y.}\ \bibnamefont {Guslienko}}, \bibinfo {author} {\bibfnamefont {V.}~\bibnamefont {Novosad}}, \bibinfo {author} {\bibfnamefont {Y.}~\bibnamefont {Otani}}, \bibinfo {author} {\bibfnamefont {H.}~\bibnamefont {Shima}},\ and\ \bibinfo {author} {\bibfnamefont {K.}~\bibnamefont {Fukamichi}},\ }\bibfield  {title} {\bibinfo {title} {Field evolution of magnetic vortex state in ferromagnetic disks},\ }\href {https://doi.org/10.1063/1.1377850} {\bibfield  {journal} {\bibinfo  {journal} {Applied Physics Letters}\ }\textbf {\bibinfo {volume} {78}},\ \bibinfo {pages} {3848} (\bibinfo {year} {2001})}\BibitemShut {NoStop}%
\bibitem [{\citenamefont {Höllinger}\ \emph {et~al.}(2003)\citenamefont {Höllinger}, \citenamefont {Killinger},\ and\ \citenamefont {Krey}}]{HOLLINGER2003178}%
  \BibitemOpen
  \bibfield  {author} {\bibinfo {author} {\bibfnamefont {R.}~\bibnamefont {Höllinger}}, \bibinfo {author} {\bibfnamefont {A.}~\bibnamefont {Killinger}},\ and\ \bibinfo {author} {\bibfnamefont {U.}~\bibnamefont {Krey}},\ }\bibfield  {title} {\bibinfo {title} {Statics and fast dynamics of nanomagnets with vortex structure},\ }\href {https://doi.org/10.1016/S0304-8853(02)01471-3} {\bibfield  {journal} {\bibinfo  {journal} {Journal of Magnetism and Magnetic Materials}\ }\textbf {\bibinfo {volume} {261}},\ \bibinfo {pages} {178} (\bibinfo {year} {2003})}\BibitemShut {NoStop}%
\bibitem [{\citenamefont {Komineas}(2006)}]{Komineas_2006}%
  \BibitemOpen
  \bibfield  {author} {\bibinfo {author} {\bibfnamefont {S.}~\bibnamefont {Komineas}},\ }\bibfield  {title} {\bibinfo {title} {A virial theorem for vortices in ferromagnetic elements},\ }\href {https://doi.org/10.1088/0305-4470/39/20/003} {\bibfield  {journal} {\bibinfo  {journal} {Journal of Physics A: Mathematical and General}\ }\textbf {\bibinfo {volume} {39}},\ \bibinfo {pages} {5669} (\bibinfo {year} {2006})}\BibitemShut {NoStop}%
\bibitem [{\citenamefont {Huber~Jr.}\ \emph {et~al.}(1958)\citenamefont {Huber~Jr.}, \citenamefont {Smith},\ and\ \citenamefont {Goodenough}}]{HuberSmithGoodenough1958}%
  \BibitemOpen
  \bibfield  {author} {\bibinfo {author} {\bibfnamefont {E.~E.}\ \bibnamefont {Huber~Jr.}}, \bibinfo {author} {\bibfnamefont {D.~O.}\ \bibnamefont {Smith}},\ and\ \bibinfo {author} {\bibfnamefont {J.~B.}\ \bibnamefont {Goodenough}},\ }\bibfield  {title} {\bibinfo {title} {Domain‐wall structure in permalloy films},\ }\href {https://doi.org/10.1063/1.1723105} {\bibfield  {journal} {\bibinfo  {journal} {Journal of Applied Physics}\ }\textbf {\bibinfo {volume} {29}},\ \bibinfo {pages} {294} (\bibinfo {year} {1958})}\BibitemShut {NoStop}%
\bibitem [{\citenamefont {Feldtkeller}\ and\ \citenamefont {Thomas}(1965)}]{Feldtkeller1965}%
  \BibitemOpen
  \bibfield  {author} {\bibinfo {author} {\bibfnamefont {E.}~\bibnamefont {Feldtkeller}}\ and\ \bibinfo {author} {\bibfnamefont {H.}~\bibnamefont {Thomas}},\ }\bibfield  {title} {\bibinfo {title} {Struktur und energie von blochlinien in d{\"u}nnen ferromagnetischen schichten},\ }\href {https://doi.org/10.1007/BF02423256} {\bibfield  {journal} {\bibinfo  {journal} {Physik der kondensierten Materie}\ }\textbf {\bibinfo {volume} {4}},\ \bibinfo {pages} {8} (\bibinfo {year} {1965})}\BibitemShut {NoStop}%
\bibitem [{\citenamefont {Nakatani}\ \emph {et~al.}(1989)\citenamefont {Nakatani}, \citenamefont {Uesaka},\ and\ \citenamefont {Hayashi}}]{Nakatani_1989}%
  \BibitemOpen
  \bibfield  {author} {\bibinfo {author} {\bibfnamefont {Y.}~\bibnamefont {Nakatani}}, \bibinfo {author} {\bibfnamefont {Y.}~\bibnamefont {Uesaka}},\ and\ \bibinfo {author} {\bibfnamefont {N.}~\bibnamefont {Hayashi}},\ }\bibfield  {title} {\bibinfo {title} {Direct solution of the {Landau-Lifshitz-Gilbert} equation for micromagnetics},\ }\href {https://doi.org/10.1143/JJAP.28.2485} {\bibfield  {journal} {\bibinfo  {journal} {Japanese Journal of Applied Physics}\ }\textbf {\bibinfo {volume} {28}},\ \bibinfo {pages} {2485} (\bibinfo {year} {1989})}\BibitemShut {NoStop}%
\bibitem [{\citenamefont {Komineas}(2007)}]{PhysRevLett.99.117202}%
  \BibitemOpen
  \bibfield  {author} {\bibinfo {author} {\bibfnamefont {S.}~\bibnamefont {Komineas}},\ }\bibfield  {title} {\bibinfo {title} {Rotating vortex dipoles in ferromagnets},\ }\href {https://doi.org/10.1103/PhysRevLett.99.117202} {\bibfield  {journal} {\bibinfo  {journal} {Phys. Rev. Lett.}\ }\textbf {\bibinfo {volume} {99}},\ \bibinfo {pages} {117202} (\bibinfo {year} {2007})}\BibitemShut {NoStop}%
\bibitem [{\citenamefont {Pereiro}\ \emph {et~al.}(2014)\citenamefont {Pereiro}, \citenamefont {Yudin}, \citenamefont {Chico}, \citenamefont {Etz}, \citenamefont {Eriksson},\ and\ \citenamefont {Bergman}}]{Pereiro2014}%
  \BibitemOpen
  \bibfield  {author} {\bibinfo {author} {\bibfnamefont {M.}~\bibnamefont {Pereiro}}, \bibinfo {author} {\bibfnamefont {D.}~\bibnamefont {Yudin}}, \bibinfo {author} {\bibfnamefont {J.}~\bibnamefont {Chico}}, \bibinfo {author} {\bibfnamefont {C.}~\bibnamefont {Etz}}, \bibinfo {author} {\bibfnamefont {O.}~\bibnamefont {Eriksson}},\ and\ \bibinfo {author} {\bibfnamefont {A.}~\bibnamefont {Bergman}},\ }\bibfield  {title} {\bibinfo {title} {Topological excitations in a kagome magnet},\ }\href {https://doi.org/10.1038/ncomms5815} {\bibfield  {journal} {\bibinfo  {journal} {Nature Communications}\ }\textbf {\bibinfo {volume} {5}},\ \bibinfo {pages} {4815} (\bibinfo {year} {2014})}\BibitemShut {NoStop}%
\bibitem [{\citenamefont {Lin}\ \emph {et~al.}(2015)\citenamefont {Lin}, \citenamefont {Saxena},\ and\ \citenamefont {Batista}}]{PhysRevB.91.224407}%
  \BibitemOpen
  \bibfield  {author} {\bibinfo {author} {\bibfnamefont {S.-Z.}\ \bibnamefont {Lin}}, \bibinfo {author} {\bibfnamefont {A.}~\bibnamefont {Saxena}},\ and\ \bibinfo {author} {\bibfnamefont {C.~D.}\ \bibnamefont {Batista}},\ }\bibfield  {title} {\bibinfo {title} {Skyrmion fractionalization and merons in chiral magnets with easy-plane anisotropy},\ }\href {https://doi.org/10.1103/PhysRevB.91.224407} {\bibfield  {journal} {\bibinfo  {journal} {Phys. Rev. B}\ }\textbf {\bibinfo {volume} {91}},\ \bibinfo {pages} {224407} (\bibinfo {year} {2015})}\BibitemShut {NoStop}%
\bibitem [{\citenamefont {Yu}\ \emph {et~al.}(2018)\citenamefont {Yu}, \citenamefont {Koshibae}, \citenamefont {Tokunaga}, \citenamefont {Shibata}, \citenamefont {Taguchi}, \citenamefont {Nagaosa},\ and\ \citenamefont {Tokura}}]{Yu2018}%
  \BibitemOpen
  \bibfield  {author} {\bibinfo {author} {\bibfnamefont {X.~Z.}\ \bibnamefont {Yu}}, \bibinfo {author} {\bibfnamefont {W.}~\bibnamefont {Koshibae}}, \bibinfo {author} {\bibfnamefont {Y.}~\bibnamefont {Tokunaga}}, \bibinfo {author} {\bibfnamefont {K.}~\bibnamefont {Shibata}}, \bibinfo {author} {\bibfnamefont {Y.}~\bibnamefont {Taguchi}}, \bibinfo {author} {\bibfnamefont {N.}~\bibnamefont {Nagaosa}},\ and\ \bibinfo {author} {\bibfnamefont {Y.}~\bibnamefont {Tokura}},\ }\bibfield  {title} {\bibinfo {title} {Transformation between meron and skyrmion topological spin textures in a chiral magnet},\ }\href {https://doi.org/10.1038/s41586-018-0745-3} {\bibfield  {journal} {\bibinfo  {journal} {Nature}\ }\textbf {\bibinfo {volume} {564}},\ \bibinfo {pages} {95} (\bibinfo {year} {2018})}\BibitemShut {NoStop}%
\bibitem [{\citenamefont {Kuchkin}\ and\ \citenamefont {Kiselev}(2020)}]{PhysRevB.101.064408}%
  \BibitemOpen
  \bibfield  {author} {\bibinfo {author} {\bibfnamefont {V.~M.}\ \bibnamefont {Kuchkin}}\ and\ \bibinfo {author} {\bibfnamefont {N.~S.}\ \bibnamefont {Kiselev}},\ }\bibfield  {title} {\bibinfo {title} {Turning a chiral skyrmion inside out},\ }\href {https://doi.org/10.1103/PhysRevB.101.064408} {\bibfield  {journal} {\bibinfo  {journal} {Phys. Rev. B}\ }\textbf {\bibinfo {volume} {101}},\ \bibinfo {pages} {064408} (\bibinfo {year} {2020})}\BibitemShut {NoStop}%
\bibitem [{\citenamefont {Zhu}\ \emph {et~al.}(2000)\citenamefont {Zhu}, \citenamefont {Zheng},\ and\ \citenamefont {Prinz}}]{Zhu2000}%
  \BibitemOpen
  \bibfield  {author} {\bibinfo {author} {\bibfnamefont {J.-G.}\ \bibnamefont {Zhu}}, \bibinfo {author} {\bibfnamefont {Y.}~\bibnamefont {Zheng}},\ and\ \bibinfo {author} {\bibfnamefont {G.~A.}\ \bibnamefont {Prinz}},\ }\bibfield  {title} {\bibinfo {title} {Ultrahigh density vertical magnetoresistive random access memory (invited)},\ }\href {https://doi.org/10.1063/1.372805} {\bibfield  {journal} {\bibinfo  {journal} {Journal of Applied Physics}\ }\textbf {\bibinfo {volume} {87}},\ \bibinfo {pages} {6668} (\bibinfo {year} {2000})}\BibitemShut {NoStop}%
\bibitem [{\citenamefont {Li}\ \emph {et~al.}(2001)\citenamefont {Li}, \citenamefont {Peyrade}, \citenamefont {Natali}, \citenamefont {Lebib}, \citenamefont {Chen}, \citenamefont {Ebels}, \citenamefont {Buda},\ and\ \citenamefont {Ounadjela}}]{PhysRevLett.86.1102}%
  \BibitemOpen
  \bibfield  {author} {\bibinfo {author} {\bibfnamefont {S.~P.}\ \bibnamefont {Li}}, \bibinfo {author} {\bibfnamefont {D.}~\bibnamefont {Peyrade}}, \bibinfo {author} {\bibfnamefont {M.}~\bibnamefont {Natali}}, \bibinfo {author} {\bibfnamefont {A.}~\bibnamefont {Lebib}}, \bibinfo {author} {\bibfnamefont {Y.}~\bibnamefont {Chen}}, \bibinfo {author} {\bibfnamefont {U.}~\bibnamefont {Ebels}}, \bibinfo {author} {\bibfnamefont {L.~D.}\ \bibnamefont {Buda}},\ and\ \bibinfo {author} {\bibfnamefont {K.}~\bibnamefont {Ounadjela}},\ }\bibfield  {title} {\bibinfo {title} {Flux closure structures in cobalt rings},\ }\href {https://doi.org/10.1103/PhysRevLett.86.1102} {\bibfield  {journal} {\bibinfo  {journal} {Phys. Rev. Lett.}\ }\textbf {\bibinfo {volume} {86}},\ \bibinfo {pages} {1102} (\bibinfo {year} {2001})}\BibitemShut {NoStop}%
\bibitem [{\citenamefont {Kläui}\ \emph {et~al.}(2003)\citenamefont {Kläui}, \citenamefont {Vaz}, \citenamefont {Lopez-Diaz},\ and\ \citenamefont {Bland}}]{Klaui_2003}%
  \BibitemOpen
  \bibfield  {author} {\bibinfo {author} {\bibfnamefont {M.}~\bibnamefont {Kläui}}, \bibinfo {author} {\bibfnamefont {C.~A.~F.}\ \bibnamefont {Vaz}}, \bibinfo {author} {\bibfnamefont {L.}~\bibnamefont {Lopez-Diaz}},\ and\ \bibinfo {author} {\bibfnamefont {J.~A.~C.}\ \bibnamefont {Bland}},\ }\bibfield  {title} {\bibinfo {title} {Vortex formation in narrow ferromagnetic rings},\ }\href {https://doi.org/10.1088/0953-8984/15/21/201} {\bibfield  {journal} {\bibinfo  {journal} {Journal of Physics: Condensed Matter}\ }\textbf {\bibinfo {volume} {15}},\ \bibinfo {pages} {R985} (\bibinfo {year} {2003})}\BibitemShut {NoStop}%
\bibitem [{\citenamefont {Muratov}\ and\ \citenamefont {Osipov}(2009)}]{MuratovOsipov2009}%
  \BibitemOpen
  \bibfield  {author} {\bibinfo {author} {\bibfnamefont {C.~B.}\ \bibnamefont {Muratov}}\ and\ \bibinfo {author} {\bibfnamefont {V.~V.}\ \bibnamefont {Osipov}},\ }\bibfield  {title} {\bibinfo {title} {Bit storage by 360$^{\circ}$ domain walls in ferromagnetic nanorings},\ }\href {https://doi.org/10.1109/TMAG.2009.2020329} {\bibfield  {journal} {\bibinfo  {journal} {IEEE Transactions on Magnetics}\ }\textbf {\bibinfo {volume} {45}},\ \bibinfo {pages} {3207} (\bibinfo {year} {2009})}\BibitemShut {NoStop}%
\bibitem [{\citenamefont {Cowburn}\ and\ \citenamefont {Welland}(1998)}]{PhysRevB.58.9217}%
  \BibitemOpen
  \bibfield  {author} {\bibinfo {author} {\bibfnamefont {R.~P.}\ \bibnamefont {Cowburn}}\ and\ \bibinfo {author} {\bibfnamefont {M.~E.}\ \bibnamefont {Welland}},\ }\bibfield  {title} {\bibinfo {title} {Micromagnetics of the single-domain state of square ferromagnetic nanostructures},\ }\href {https://doi.org/10.1103/PhysRevB.58.9217} {\bibfield  {journal} {\bibinfo  {journal} {Phys. Rev. B}\ }\textbf {\bibinfo {volume} {58}},\ \bibinfo {pages} {9217} (\bibinfo {year} {1998})}\BibitemShut {NoStop}%
\bibitem [{\citenamefont {Moser}(2004)}]{Moser2004}%
  \BibitemOpen
  \bibfield  {author} {\bibinfo {author} {\bibfnamefont {R.}~\bibnamefont {Moser}},\ }\bibfield  {title} {\bibinfo {title} {Boundary vortices for thin ferromagnetic films},\ }\href {https://doi.org/10.1007/s00205-004-0329-2} {\bibfield  {journal} {\bibinfo  {journal} {Archive for Rational Mechanics and Analysis}\ }\textbf {\bibinfo {volume} {174}},\ \bibinfo {pages} {267} (\bibinfo {year} {2004})}\BibitemShut {NoStop}%
\bibitem [{\citenamefont {Kravchuk}\ \emph {et~al.}(2012)\citenamefont {Kravchuk}, \citenamefont {Sheka}, \citenamefont {Streubel}, \citenamefont {Makarov}, \citenamefont {Schmidt},\ and\ \citenamefont {Gaididei}}]{PhysRevB.85.144433}%
  \BibitemOpen
  \bibfield  {author} {\bibinfo {author} {\bibfnamefont {V.~P.}\ \bibnamefont {Kravchuk}}, \bibinfo {author} {\bibfnamefont {D.~D.}\ \bibnamefont {Sheka}}, \bibinfo {author} {\bibfnamefont {R.}~\bibnamefont {Streubel}}, \bibinfo {author} {\bibfnamefont {D.}~\bibnamefont {Makarov}}, \bibinfo {author} {\bibfnamefont {O.~G.}\ \bibnamefont {Schmidt}},\ and\ \bibinfo {author} {\bibfnamefont {Y.}~\bibnamefont {Gaididei}},\ }\bibfield  {title} {\bibinfo {title} {Out-of-surface vortices in spherical shells},\ }\href {https://doi.org/10.1103/PhysRevB.85.144433} {\bibfield  {journal} {\bibinfo  {journal} {Phys. Rev. B}\ }\textbf {\bibinfo {volume} {85}},\ \bibinfo {pages} {144433} (\bibinfo {year} {2012})}\BibitemShut {NoStop}%
\bibitem [{\citenamefont {Streubel}\ \emph {et~al.}(2016)\citenamefont {Streubel}, \citenamefont {Fischer}, \citenamefont {Kronast}, \citenamefont {Kravchuk}, \citenamefont {Sheka}, \citenamefont {Gaididei}, \citenamefont {Schmidt},\ and\ \citenamefont {Makarov}}]{Streubel_2016}%
  \BibitemOpen
  \bibfield  {author} {\bibinfo {author} {\bibfnamefont {R.}~\bibnamefont {Streubel}}, \bibinfo {author} {\bibfnamefont {P.}~\bibnamefont {Fischer}}, \bibinfo {author} {\bibfnamefont {F.}~\bibnamefont {Kronast}}, \bibinfo {author} {\bibfnamefont {V.~P.}\ \bibnamefont {Kravchuk}}, \bibinfo {author} {\bibfnamefont {D.~D.}\ \bibnamefont {Sheka}}, \bibinfo {author} {\bibfnamefont {Y.}~\bibnamefont {Gaididei}}, \bibinfo {author} {\bibfnamefont {O.~G.}\ \bibnamefont {Schmidt}},\ and\ \bibinfo {author} {\bibfnamefont {D.}~\bibnamefont {Makarov}},\ }\bibfield  {title} {\bibinfo {title} {Magnetism in curved geometries},\ }\href {https://doi.org/10.1088/0022-3727/49/36/363001} {\bibfield  {journal} {\bibinfo  {journal} {Journal of Physics D}\ }\textbf {\bibinfo {volume} {49}},\ \bibinfo {pages} {363001} (\bibinfo {year} {2016})}\BibitemShut {NoStop}%
\bibitem [{\citenamefont {Volkov}\ \emph {et~al.}(2024)\citenamefont {Volkov}, \citenamefont {Pylypovskyi}, \citenamefont {Porrati}, \citenamefont {Kronast}, \citenamefont {Fernandez-Roldan}, \citenamefont {K{\'a}kay}, \citenamefont {Kuprava}, \citenamefont {Barth}, \citenamefont {Rybakov}, \citenamefont {Eriksson}, \citenamefont {Lamb-Camarena}, \citenamefont {Makushko}, \citenamefont {Mawass}, \citenamefont {Shakeel}, \citenamefont {Dobrovolskiy}, \citenamefont {Huth},\ and\ \citenamefont {Makarov}}]{Volkov2024}%
  \BibitemOpen
  \bibfield  {author} {\bibinfo {author} {\bibfnamefont {O.~M.}\ \bibnamefont {Volkov}}, \bibinfo {author} {\bibfnamefont {O.~V.}\ \bibnamefont {Pylypovskyi}}, \bibinfo {author} {\bibfnamefont {F.}~\bibnamefont {Porrati}}, \bibinfo {author} {\bibfnamefont {F.}~\bibnamefont {Kronast}}, \bibinfo {author} {\bibfnamefont {J.~A.}\ \bibnamefont {Fernandez-Roldan}}, \bibinfo {author} {\bibfnamefont {A.}~\bibnamefont {K{\'a}kay}}, \bibinfo {author} {\bibfnamefont {A.}~\bibnamefont {Kuprava}}, \bibinfo {author} {\bibfnamefont {S.}~\bibnamefont {Barth}}, \bibinfo {author} {\bibfnamefont {F.~N.}\ \bibnamefont {Rybakov}}, \bibinfo {author} {\bibfnamefont {O.}~\bibnamefont {Eriksson}}, \bibinfo {author} {\bibfnamefont {S.}~\bibnamefont {Lamb-Camarena}}, \bibinfo {author} {\bibfnamefont {P.}~\bibnamefont {Makushko}}, \bibinfo {author} {\bibfnamefont {M.-A.}\ \bibnamefont {Mawass}}, \bibinfo {author} {\bibfnamefont {S.}~\bibnamefont {Shakeel}}, \bibinfo {author} {\bibfnamefont {O.~V.}\ \bibnamefont {Dobrovolskiy}}, \bibinfo
  {author} {\bibfnamefont {M.}~\bibnamefont {Huth}},\ and\ \bibinfo {author} {\bibfnamefont {D.}~\bibnamefont {Makarov}},\ }\bibfield  {title} {\bibinfo {title} {Three-dimensional magnetic nanotextures with high-order vorticity in soft magnetic wireframes},\ }\href {https://doi.org/10.1038/s41467-024-46403-8} {\bibfield  {journal} {\bibinfo  {journal} {Nature Communications}\ }\textbf {\bibinfo {volume} {15}},\ \bibinfo {pages} {2193} (\bibinfo {year} {2024})}\BibitemShut {NoStop}%
\bibitem [{\citenamefont {Berezinskii}(1972)}]{Berezinskii_1972}%
  \BibitemOpen
  \bibfield  {author} {\bibinfo {author} {\bibfnamefont {V.~L.}\ \bibnamefont {Berezinskii}},\ }\bibfield  {title} {\bibinfo {title} {Destruction of long-range order in one-dimensional and two-dimensional systems possessing a continuous symmetry group. {II}. quantum systems},\ }\href {http://www.jetp.ras.ru/cgi-bin/e/index/r/61/3/p1144?a=list} {\bibfield  {journal} {\bibinfo  {journal} {Soviet Physics JETP}\ }\textbf {\bibinfo {volume} {34}},\ \bibinfo {pages} {610} (\bibinfo {year} {1972})}\BibitemShut {NoStop}%
\bibitem [{\citenamefont {Kosterlitz}\ and\ \citenamefont {Thouless}(1973)}]{Kosterlitz_1973}%
  \BibitemOpen
  \bibfield  {author} {\bibinfo {author} {\bibfnamefont {J.~M.}\ \bibnamefont {Kosterlitz}}\ and\ \bibinfo {author} {\bibfnamefont {D.~J.}\ \bibnamefont {Thouless}},\ }\bibfield  {title} {\bibinfo {title} {Ordering, metastability and phase transitions in two-dimensional systems},\ }\href {https://doi.org/10.1088/0022-3719/6/7/010} {\bibfield  {journal} {\bibinfo  {journal} {Journal of Physics C}\ }\textbf {\bibinfo {volume} {6}},\ \bibinfo {pages} {1181} (\bibinfo {year} {1973})}\BibitemShut {NoStop}%
\bibitem [{\citenamefont {José}(2013)}]{FourtyYearsBKT}%
  \BibitemOpen
  \bibinfo {editor} {\bibfnamefont {J.~V.}\ \bibnamefont {José}},\ ed.,\ \href {https://doi.org/10.1142/8572} {\emph {\bibinfo {title} {40 Years of {Berezinskii}-{Kosterlitz}-{Thouless} Theory}}}\ (\bibinfo  {publisher} {World Scientific},\ \bibinfo {year} {2013})\BibitemShut {NoStop}%
\bibitem [{\citenamefont {Svistunov}\ \emph {et~al.}(2015)\citenamefont {Svistunov}, \citenamefont {Babaev},\ and\ \citenamefont {Prokof'ev}}]{SvistunovBabaevProkofev}%
  \BibitemOpen
  \bibfield  {author} {\bibinfo {author} {\bibfnamefont {B.}~\bibnamefont {Svistunov}}, \bibinfo {author} {\bibfnamefont {E.}~\bibnamefont {Babaev}},\ and\ \bibinfo {author} {\bibfnamefont {N.}~\bibnamefont {Prokof'ev}},\ }\href {https://doi.org/10.1201/b18346} {\emph {\bibinfo {title} {Superfluid states of matter}}}\ (\bibinfo  {publisher} {CRC Press},\ \bibinfo {year} {2015})\BibitemShut {NoStop}%
\bibitem [{\citenamefont {Senthil}\ \emph {et~al.}(2004)\citenamefont {Senthil}, \citenamefont {Vishwanath}, \citenamefont {Balents}, \citenamefont {Sachdev},\ and\ \citenamefont {Fisher}}]{Senthil_2004}%
  \BibitemOpen
  \bibfield  {author} {\bibinfo {author} {\bibfnamefont {T.}~\bibnamefont {Senthil}}, \bibinfo {author} {\bibfnamefont {A.}~\bibnamefont {Vishwanath}}, \bibinfo {author} {\bibfnamefont {L.}~\bibnamefont {Balents}}, \bibinfo {author} {\bibfnamefont {S.}~\bibnamefont {Sachdev}},\ and\ \bibinfo {author} {\bibfnamefont {M.~P.~A.}\ \bibnamefont {Fisher}},\ }\bibfield  {title} {\bibinfo {title} {Deconfined quantum critical points},\ }\href {https://doi.org/10.1126/science.1091806} {\bibfield  {journal} {\bibinfo  {journal} {Science}\ }\textbf {\bibinfo {volume} {303}},\ \bibinfo {pages} {1490} (\bibinfo {year} {2004})}\BibitemShut {NoStop}%
\bibitem [{\citenamefont {Göbel}\ \emph {et~al.}(2021)\citenamefont {Göbel}, \citenamefont {Mertig},\ and\ \citenamefont {Tretiakov}}]{GOBEL20211}%
  \BibitemOpen
  \bibfield  {author} {\bibinfo {author} {\bibfnamefont {B.}~\bibnamefont {Göbel}}, \bibinfo {author} {\bibfnamefont {I.}~\bibnamefont {Mertig}},\ and\ \bibinfo {author} {\bibfnamefont {O.~A.}\ \bibnamefont {Tretiakov}},\ }\bibfield  {title} {\bibinfo {title} {Beyond skyrmions: Review and perspectives of alternative magnetic quasiparticles},\ }\href {https://doi.org/10.1016/j.physrep.2020.10.001} {\bibfield  {journal} {\bibinfo  {journal} {Physics Reports}\ }\textbf {\bibinfo {volume} {895}},\ \bibinfo {pages} {1} (\bibinfo {year} {2021})}\BibitemShut {NoStop}%
\bibitem [{\citenamefont {Augustin}\ \emph {et~al.}(2021)\citenamefont {Augustin}, \citenamefont {Jenkins}, \citenamefont {Evans}, \citenamefont {Novoselov},\ and\ \citenamefont {Santos}}]{Augustin2021}%
  \BibitemOpen
  \bibfield  {author} {\bibinfo {author} {\bibfnamefont {M.}~\bibnamefont {Augustin}}, \bibinfo {author} {\bibfnamefont {S.}~\bibnamefont {Jenkins}}, \bibinfo {author} {\bibfnamefont {R.~F.~L.}\ \bibnamefont {Evans}}, \bibinfo {author} {\bibfnamefont {K.~S.}\ \bibnamefont {Novoselov}},\ and\ \bibinfo {author} {\bibfnamefont {E.~J.~G.}\ \bibnamefont {Santos}},\ }\bibfield  {title} {\bibinfo {title} {Properties and dynamics of meron topological spin textures in the two-dimensional magnet {CrCl$_3$}},\ }\href {https://doi.org/10.1038/s41467-020-20497-2} {\bibfield  {journal} {\bibinfo  {journal} {Nature Communications}\ }\textbf {\bibinfo {volume} {12}},\ \bibinfo {pages} {185} (\bibinfo {year} {2021})}\BibitemShut {NoStop}%
\bibitem [{\citenamefont {Strungaru}\ \emph {et~al.}(2022)\citenamefont {Strungaru}, \citenamefont {Augustin},\ and\ \citenamefont {Santos}}]{Strungaru2022}%
  \BibitemOpen
  \bibfield  {author} {\bibinfo {author} {\bibfnamefont {M.}~\bibnamefont {Strungaru}}, \bibinfo {author} {\bibfnamefont {M.}~\bibnamefont {Augustin}},\ and\ \bibinfo {author} {\bibfnamefont {E.~J.~G.}\ \bibnamefont {Santos}},\ }\bibfield  {title} {\bibinfo {title} {Ultrafast laser-driven topological spin textures on a {2D} magnet},\ }\href {https://doi.org/10.1038/s41524-022-00864-x} {\bibfield  {journal} {\bibinfo  {journal} {npj Computational Materials}\ }\textbf {\bibinfo {volume} {8}},\ \bibinfo {pages} {169} (\bibinfo {year} {2022})}\BibitemShut {NoStop}%
\bibitem [{\citenamefont {Allan}\ \emph {et~al.}(1977)\citenamefont {Allan}, \citenamefont {Dashen},\ and\ \citenamefont {Gross}}]{ALLAN1977375}%
  \BibitemOpen
  \bibfield  {author} {\bibinfo {author} {\bibfnamefont {C.~G.}\ \bibnamefont {Allan}}, \bibinfo {author} {\bibfnamefont {R.}~\bibnamefont {Dashen}},\ and\ \bibinfo {author} {\bibfnamefont {D.~J.}\ \bibnamefont {Gross}},\ }\bibfield  {title} {\bibinfo {title} {A mechanism for quark confinement},\ }\href {https://doi.org/10.1016/0370-2693(77)90019-3} {\bibfield  {journal} {\bibinfo  {journal} {Physics Letters B}\ }\textbf {\bibinfo {volume} {66}},\ \bibinfo {pages} {375} (\bibinfo {year} {1977})}\BibitemShut {NoStop}%
\bibitem [{\citenamefont {Gross}(1978)}]{GROSS1978439}%
  \BibitemOpen
  \bibfield  {author} {\bibinfo {author} {\bibfnamefont {D.~J.}\ \bibnamefont {Gross}},\ }\bibfield  {title} {\bibinfo {title} {Meron configurations in the two-dimensional {$O(3)$} $\sigma$-model},\ }\href {https://doi.org/10.1016/0550-3213(78)90470-4} {\bibfield  {journal} {\bibinfo  {journal} {Nuclear Physics B}\ }\textbf {\bibinfo {volume} {132}},\ \bibinfo {pages} {439} (\bibinfo {year} {1978})}\BibitemShut {NoStop}%
\bibitem [{\citenamefont {Volovik}(2003)}]{Volovik_book}%
  \BibitemOpen
  \bibfield  {author} {\bibinfo {author} {\bibfnamefont {G.~E.}\ \bibnamefont {Volovik}},\ }\href {https://doi.org/10.1093/acprof:oso/9780199564842.001.0001} {\emph {\bibinfo {title} {The Universe in a Helium Droplet}}},\ International Series of Monographs on Physics, Vol. 117\ (\bibinfo  {publisher} {Oxford University Press},\ \bibinfo {year} {2003})\BibitemShut {NoStop}%
\bibitem [{\citenamefont {Van~Waeyenberge}\ \emph {et~al.}(2006)\citenamefont {Van~Waeyenberge}, \citenamefont {Puzic}, \citenamefont {Stoll}, \citenamefont {Chou}, \citenamefont {Tyliszczak}, \citenamefont {Hertel}, \citenamefont {F{\"a}hnle}, \citenamefont {Br{\"u}ckl}, \citenamefont {Rott}, \citenamefont {Reiss}, \citenamefont {Neudecker}, \citenamefont {Weiss}, \citenamefont {Back},\ and\ \citenamefont {Sch{\"u}tz}}]{VanWaeyenberge2006}%
  \BibitemOpen
  \bibfield  {author} {\bibinfo {author} {\bibfnamefont {B.}~\bibnamefont {Van~Waeyenberge}}, \bibinfo {author} {\bibfnamefont {A.}~\bibnamefont {Puzic}}, \bibinfo {author} {\bibfnamefont {H.}~\bibnamefont {Stoll}}, \bibinfo {author} {\bibfnamefont {K.~W.}\ \bibnamefont {Chou}}, \bibinfo {author} {\bibfnamefont {T.}~\bibnamefont {Tyliszczak}}, \bibinfo {author} {\bibfnamefont {R.}~\bibnamefont {Hertel}}, \bibinfo {author} {\bibfnamefont {M.}~\bibnamefont {F{\"a}hnle}}, \bibinfo {author} {\bibfnamefont {H.}~\bibnamefont {Br{\"u}ckl}}, \bibinfo {author} {\bibfnamefont {K.}~\bibnamefont {Rott}}, \bibinfo {author} {\bibfnamefont {G.}~\bibnamefont {Reiss}}, \bibinfo {author} {\bibfnamefont {I.}~\bibnamefont {Neudecker}}, \bibinfo {author} {\bibfnamefont {D.}~\bibnamefont {Weiss}}, \bibinfo {author} {\bibfnamefont {C.~H.}\ \bibnamefont {Back}},\ and\ \bibinfo {author} {\bibfnamefont {G.}~\bibnamefont {Sch{\"u}tz}},\ }\bibfield  {title} {\bibinfo {title} {Magnetic vortex core reversal by excitation with short bursts
  of an alternating field},\ }\href {https://doi.org/10.1038/nature05240} {\bibfield  {journal} {\bibinfo  {journal} {Nature}\ }\textbf {\bibinfo {volume} {444}},\ \bibinfo {pages} {461} (\bibinfo {year} {2006})}\BibitemShut {NoStop}%
\bibitem [{\citenamefont {Tretiakov}\ and\ \citenamefont {Tchernyshyov}(2007)}]{PhysRevB.75.012408}%
  \BibitemOpen
  \bibfield  {author} {\bibinfo {author} {\bibfnamefont {O.~A.}\ \bibnamefont {Tretiakov}}\ and\ \bibinfo {author} {\bibfnamefont {O.}~\bibnamefont {Tchernyshyov}},\ }\bibfield  {title} {\bibinfo {title} {Vortices in thin ferromagnetic films and the skyrmion number},\ }\href {https://doi.org/10.1103/PhysRevB.75.012408} {\bibfield  {journal} {\bibinfo  {journal} {Phys. Rev. B}\ }\textbf {\bibinfo {volume} {75}},\ \bibinfo {pages} {012408} (\bibinfo {year} {2007})}\BibitemShut {NoStop}%
\bibitem [{\citenamefont {Hatcher}(2002)}]{Hatcher}%
  \BibitemOpen
  \bibfield  {author} {\bibinfo {author} {\bibfnamefont {A.}~\bibnamefont {Hatcher}},\ }\href@noop {} {\emph {\bibinfo {title} {Algebraic Topology}}}\ (\bibinfo  {publisher} {Cambridge University Press},\ \bibinfo {year} {2002})\BibitemShut {NoStop}%
\bibitem [{\citenamefont {Zhang}\ \emph {et~al.}(2015)\citenamefont {Zhang}, \citenamefont {Ezawa},\ and\ \citenamefont {Zhou}}]{Zhang2015}%
  \BibitemOpen
  \bibfield  {author} {\bibinfo {author} {\bibfnamefont {X.}~\bibnamefont {Zhang}}, \bibinfo {author} {\bibfnamefont {M.}~\bibnamefont {Ezawa}},\ and\ \bibinfo {author} {\bibfnamefont {Y.}~\bibnamefont {Zhou}},\ }\bibfield  {title} {\bibinfo {title} {Magnetic skyrmion logic gates: conversion, duplication and merging of skyrmions},\ }\href {https://doi.org/10.1038/srep09400} {\bibfield  {journal} {\bibinfo  {journal} {Scientific Reports}\ }\textbf {\bibinfo {volume} {5}},\ \bibinfo {pages} {9400} (\bibinfo {year} {2015})}\BibitemShut {NoStop}%
\bibitem [{\citenamefont {G\"obel}\ \emph {et~al.}(2019)\citenamefont {G\"obel}, \citenamefont {Mook}, \citenamefont {Henk}, \citenamefont {Mertig},\ and\ \citenamefont {Tretiakov}}]{PhysRevB.99.060407}%
  \BibitemOpen
  \bibfield  {author} {\bibinfo {author} {\bibfnamefont {B.}~\bibnamefont {G\"obel}}, \bibinfo {author} {\bibfnamefont {A.}~\bibnamefont {Mook}}, \bibinfo {author} {\bibfnamefont {J.}~\bibnamefont {Henk}}, \bibinfo {author} {\bibfnamefont {I.}~\bibnamefont {Mertig}},\ and\ \bibinfo {author} {\bibfnamefont {O.~A.}\ \bibnamefont {Tretiakov}},\ }\bibfield  {title} {\bibinfo {title} {Magnetic bimerons as skyrmion analogues in in-plane magnets},\ }\href {https://doi.org/10.1103/PhysRevB.99.060407} {\bibfield  {journal} {\bibinfo  {journal} {Phys. Rev. B}\ }\textbf {\bibinfo {volume} {99}},\ \bibinfo {pages} {060407} (\bibinfo {year} {2019})}\BibitemShut {NoStop}%
\bibitem [{\citenamefont {Gao}\ \emph {et~al.}(2019)\citenamefont {Gao}, \citenamefont {Je}, \citenamefont {Im}, \citenamefont {Choi}, \citenamefont {Yang}, \citenamefont {Li}, \citenamefont {Wang}, \citenamefont {Lee}, \citenamefont {Han}, \citenamefont {Lee}, \citenamefont {Chao}, \citenamefont {Hwang}, \citenamefont {Li},\ and\ \citenamefont {Qiu}}]{Gao2019}%
  \BibitemOpen
  \bibfield  {author} {\bibinfo {author} {\bibfnamefont {N.}~\bibnamefont {Gao}}, \bibinfo {author} {\bibfnamefont {S.-G.}\ \bibnamefont {Je}}, \bibinfo {author} {\bibfnamefont {M.-Y.}\ \bibnamefont {Im}}, \bibinfo {author} {\bibfnamefont {J.~W.}\ \bibnamefont {Choi}}, \bibinfo {author} {\bibfnamefont {M.}~\bibnamefont {Yang}}, \bibinfo {author} {\bibfnamefont {Q.}~\bibnamefont {Li}}, \bibinfo {author} {\bibfnamefont {T.~Y.}\ \bibnamefont {Wang}}, \bibinfo {author} {\bibfnamefont {S.}~\bibnamefont {Lee}}, \bibinfo {author} {\bibfnamefont {H.-S.}\ \bibnamefont {Han}}, \bibinfo {author} {\bibfnamefont {K.-S.}\ \bibnamefont {Lee}}, \bibinfo {author} {\bibfnamefont {W.}~\bibnamefont {Chao}}, \bibinfo {author} {\bibfnamefont {C.}~\bibnamefont {Hwang}}, \bibinfo {author} {\bibfnamefont {J.}~\bibnamefont {Li}},\ and\ \bibinfo {author} {\bibfnamefont {Z.~Q.}\ \bibnamefont {Qiu}},\ }\bibfield  {title} {\bibinfo {title} {Creation and annihilation of topological meron pairs in in-plane magnetized films},\ }\href
  {https://doi.org/10.1038/s41467-019-13642-z} {\bibfield  {journal} {\bibinfo  {journal} {Nature Communications}\ }\textbf {\bibinfo {volume} {10}},\ \bibinfo {pages} {5603} (\bibinfo {year} {2019})}\BibitemShut {NoStop}%
\bibitem [{\citenamefont {Bachmann}\ \emph {et~al.}(2023)\citenamefont {Bachmann}, \citenamefont {Lianeris},\ and\ \citenamefont {Komineas}}]{PhysRevB.108.014402}%
  \BibitemOpen
  \bibfield  {author} {\bibinfo {author} {\bibfnamefont {D.}~\bibnamefont {Bachmann}}, \bibinfo {author} {\bibfnamefont {M.}~\bibnamefont {Lianeris}},\ and\ \bibinfo {author} {\bibfnamefont {S.}~\bibnamefont {Komineas}},\ }\bibfield  {title} {\bibinfo {title} {Meron configurations in easy-plane chiral magnets},\ }\href {https://doi.org/10.1103/PhysRevB.108.014402} {\bibfield  {journal} {\bibinfo  {journal} {Phys. Rev. B}\ }\textbf {\bibinfo {volume} {108}},\ \bibinfo {pages} {014402} (\bibinfo {year} {2023})}\BibitemShut {NoStop}%
\bibitem [{\citenamefont {J\"aykk\"a}\ and\ \citenamefont {Speight}(2010)}]{PhysRevD.82.125030}%
  \BibitemOpen
  \bibfield  {author} {\bibinfo {author} {\bibfnamefont {J.}~\bibnamefont {J\"aykk\"a}}\ and\ \bibinfo {author} {\bibfnamefont {M.}~\bibnamefont {Speight}},\ }\bibfield  {title} {\bibinfo {title} {Easy plane baby {Skyrmions}},\ }\href {https://doi.org/10.1103/PhysRevD.82.125030} {\bibfield  {journal} {\bibinfo  {journal} {Phys. Rev. D}\ }\textbf {\bibinfo {volume} {82}},\ \bibinfo {pages} {125030} (\bibinfo {year} {2010})}\BibitemShut {NoStop}%
\bibitem [{\citenamefont {Kobayashi}\ and\ \citenamefont {Nitta}(2013)}]{PhysRevD.87.125013}%
  \BibitemOpen
  \bibfield  {author} {\bibinfo {author} {\bibfnamefont {M.}~\bibnamefont {Kobayashi}}\ and\ \bibinfo {author} {\bibfnamefont {M.}~\bibnamefont {Nitta}},\ }\bibfield  {title} {\bibinfo {title} {Fractional vortex molecules and vortex polygons in a baby {Skyrme} model},\ }\href {https://doi.org/10.1103/PhysRevD.87.125013} {\bibfield  {journal} {\bibinfo  {journal} {Phys. Rev. D}\ }\textbf {\bibinfo {volume} {87}},\ \bibinfo {pages} {125013} (\bibinfo {year} {2013})}\BibitemShut {NoStop}%
\bibitem [{\citenamefont {Moon}\ \emph {et~al.}(2019)\citenamefont {Moon}, \citenamefont {Yoon}, \citenamefont {Kim},\ and\ \citenamefont {Hwang}}]{moon2019}%
  \BibitemOpen
  \bibfield  {author} {\bibinfo {author} {\bibfnamefont {K.-W.}\ \bibnamefont {Moon}}, \bibinfo {author} {\bibfnamefont {J.}~\bibnamefont {Yoon}}, \bibinfo {author} {\bibfnamefont {C.}~\bibnamefont {Kim}},\ and\ \bibinfo {author} {\bibfnamefont {C.}~\bibnamefont {Hwang}},\ }\bibfield  {title} {\bibinfo {title} {Existence of in-plane magnetic skyrmion and its motion under current flow},\ }\href {https://doi.org/10.1103/PhysRevApplied.12.064054} {\bibfield  {journal} {\bibinfo  {journal} {Phys. Rev. Appl.}\ }\textbf {\bibinfo {volume} {12}},\ \bibinfo {pages} {064054} (\bibinfo {year} {2019})}\BibitemShut {NoStop}%
\bibitem [{\citenamefont {Thiele}(1969)}]{Thiele_1969}%
  \BibitemOpen
  \bibfield  {author} {\bibinfo {author} {\bibfnamefont {A.~A.}\ \bibnamefont {Thiele}},\ }\bibfield  {title} {\bibinfo {title} {The theory of cylindrical magnetic domains},\ }\href {https://doi.org/10.1002/j.1538-7305.1969.tb01747.x} {\bibfield  {journal} {\bibinfo  {journal} {The Bell System Technical Journal}\ }\textbf {\bibinfo {volume} {48}},\ \bibinfo {pages} {3287} (\bibinfo {year} {1969})}\BibitemShut {NoStop}%
\bibitem [{\citenamefont {Bobeck}\ and\ \citenamefont {Scovil}(1971)}]{Bobeck_Scovil_1971}%
  \BibitemOpen
  \bibfield  {author} {\bibinfo {author} {\bibfnamefont {A.~H.}\ \bibnamefont {Bobeck}}\ and\ \bibinfo {author} {\bibfnamefont {H.~E.~D.}\ \bibnamefont {Scovil}},\ }\bibfield  {title} {\bibinfo {title} {Magnetic bubbles},\ }\href {https://www.jstor.org/stable/24922754} {\bibfield  {journal} {\bibinfo  {journal} {Scientific American}\ }\textbf {\bibinfo {volume} {224}},\ \bibinfo {pages} {78} (\bibinfo {year} {1971})}\BibitemShut {NoStop}%
\bibitem [{\citenamefont {Hassan}\ \emph {et~al.}(2024)\citenamefont {Hassan}, \citenamefont {Koraltan}, \citenamefont {Ullrich}, \citenamefont {Bruckner}, \citenamefont {Serha}, \citenamefont {Levchenko}, \citenamefont {Varvaro}, \citenamefont {Kiselev}, \citenamefont {Heigl}, \citenamefont {Abert}, \citenamefont {Suess},\ and\ \citenamefont {Albrecht}}]{Hassan2024}%
  \BibitemOpen
  \bibfield  {author} {\bibinfo {author} {\bibfnamefont {M.}~\bibnamefont {Hassan}}, \bibinfo {author} {\bibfnamefont {S.}~\bibnamefont {Koraltan}}, \bibinfo {author} {\bibfnamefont {A.}~\bibnamefont {Ullrich}}, \bibinfo {author} {\bibfnamefont {F.}~\bibnamefont {Bruckner}}, \bibinfo {author} {\bibfnamefont {R.~O.}\ \bibnamefont {Serha}}, \bibinfo {author} {\bibfnamefont {K.~V.}\ \bibnamefont {Levchenko}}, \bibinfo {author} {\bibfnamefont {G.}~\bibnamefont {Varvaro}}, \bibinfo {author} {\bibfnamefont {N.~S.}\ \bibnamefont {Kiselev}}, \bibinfo {author} {\bibfnamefont {M.}~\bibnamefont {Heigl}}, \bibinfo {author} {\bibfnamefont {C.}~\bibnamefont {Abert}}, \bibinfo {author} {\bibfnamefont {D.}~\bibnamefont {Suess}},\ and\ \bibinfo {author} {\bibfnamefont {M.}~\bibnamefont {Albrecht}},\ }\bibfield  {title} {\bibinfo {title} {Dipolar skyrmions and antiskyrmions of arbitrary topological charge at room temperature},\ }\href {https://doi.org/10.1038/s41567-023-02358-z} {\bibfield  {journal} {\bibinfo  {journal}
  {Nature Physics}\ }\textbf {\bibinfo {volume} {20}},\ \bibinfo {pages} {615} (\bibinfo {year} {2024})}\BibitemShut {NoStop}%
\bibitem [{\citenamefont {Kovalev}\ \emph {et~al.}(1979)\citenamefont {Kovalev}, \citenamefont {Kosevich},\ and\ \citenamefont {Maslov}}]{KovalevKosevichMaslov}%
  \BibitemOpen
  \bibfield  {author} {\bibinfo {author} {\bibfnamefont {A.~S.}\ \bibnamefont {Kovalev}}, \bibinfo {author} {\bibfnamefont {A.~M.}\ \bibnamefont {Kosevich}},\ and\ \bibinfo {author} {\bibfnamefont {K.~V.}\ \bibnamefont {Maslov}},\ }\bibfield  {title} {\bibinfo {title} {Magnetic vortex -- topological soliton in a ferromagnet with an easy-axis anisotropy},\ }\href {http://jetpletters.ru/ps/1365/article_20632.shtml} {\bibfield  {journal} {\bibinfo  {journal} {JETP Letters}\ }\textbf {\bibinfo {volume} {30}},\ \bibinfo {pages} {321} (\bibinfo {year} {1979})}\BibitemShut {NoStop}%
\bibitem [{\citenamefont {Kosevich}\ \emph {et~al.}(1990)\citenamefont {Kosevich}, \citenamefont {Ivanov},\ and\ \citenamefont {Kovalev}}]{KOSEVICH1990117}%
  \BibitemOpen
  \bibfield  {author} {\bibinfo {author} {\bibfnamefont {A.}~\bibnamefont {Kosevich}}, \bibinfo {author} {\bibfnamefont {B.}~\bibnamefont {Ivanov}},\ and\ \bibinfo {author} {\bibfnamefont {A.}~\bibnamefont {Kovalev}},\ }\bibfield  {title} {\bibinfo {title} {Magnetic solitons},\ }\href {https://doi.org/10.1016/0370-1573(90)90130-T} {\bibfield  {journal} {\bibinfo  {journal} {Physics Reports}\ }\textbf {\bibinfo {volume} {194}},\ \bibinfo {pages} {117} (\bibinfo {year} {1990})}\BibitemShut {NoStop}%
\bibitem [{\citenamefont {Bogdanov}\ and\ \citenamefont {Yablonskii}(1989)}]{Bogdanov_89}%
  \BibitemOpen
  \bibfield  {author} {\bibinfo {author} {\bibfnamefont {A.~N.}\ \bibnamefont {Bogdanov}}\ and\ \bibinfo {author} {\bibfnamefont {D.~A.}\ \bibnamefont {Yablonskii}},\ }\bibfield  {title} {\bibinfo {title} {Thermodynamically stable "vortices" in magnetically ordered crystals. {The} mixed state of magnets},\ }\href {http://www.jetp.ras.ru/cgi-bin/e/index/e/68/1/p101?a=list} {\bibfield  {journal} {\bibinfo  {journal} {Sov. Phys. JETP}\ }\textbf {\bibinfo {volume} {68}},\ \bibinfo {pages} {101} (\bibinfo {year} {1989})}\BibitemShut {NoStop}%
\bibitem [{\citenamefont {Kiselev}\ \emph {et~al.}(2011)\citenamefont {Kiselev}, \citenamefont {Bogdanov}, \citenamefont {Schäfer},\ and\ \citenamefont {Rößler}}]{Kiselev_2011}%
  \BibitemOpen
  \bibfield  {author} {\bibinfo {author} {\bibfnamefont {N.~S.}\ \bibnamefont {Kiselev}}, \bibinfo {author} {\bibfnamefont {A.~N.}\ \bibnamefont {Bogdanov}}, \bibinfo {author} {\bibfnamefont {R.}~\bibnamefont {Schäfer}},\ and\ \bibinfo {author} {\bibfnamefont {U.~K.}\ \bibnamefont {Rößler}},\ }\bibfield  {title} {\bibinfo {title} {Chiral skyrmions in thin magnetic films: new objects for magnetic storage technologies?},\ }\href {https://doi.org/10.1088/0022-3727/44/39/392001} {\bibfield  {journal} {\bibinfo  {journal} {Journal of Physics D}\ }\textbf {\bibinfo {volume} {44}},\ \bibinfo {pages} {392001} (\bibinfo {year} {2011})}\BibitemShut {NoStop}%
\bibitem [{\citenamefont {Nagaosa}\ and\ \citenamefont {Tokura}(2013)}]{Nagaosa2013}%
  \BibitemOpen
  \bibfield  {author} {\bibinfo {author} {\bibfnamefont {N.}~\bibnamefont {Nagaosa}}\ and\ \bibinfo {author} {\bibfnamefont {Y.}~\bibnamefont {Tokura}},\ }\bibfield  {title} {\bibinfo {title} {Topological properties and dynamics of magnetic skyrmions},\ }\href {https://doi.org/10.1038/nnano.2013.243} {\bibfield  {journal} {\bibinfo  {journal} {Nature Nanotechnology}\ }\textbf {\bibinfo {volume} {8}},\ \bibinfo {pages} {899} (\bibinfo {year} {2013})}\BibitemShut {NoStop}%
\bibitem [{\citenamefont {Melcher}(2014)}]{Melcher_2014}%
  \BibitemOpen
  \bibfield  {author} {\bibinfo {author} {\bibfnamefont {C.}~\bibnamefont {Melcher}},\ }\bibfield  {title} {\bibinfo {title} {Chiral skyrmions in the plane},\ }\href {https://doi.org/10.1098/rspa.2014.0394} {\bibfield  {journal} {\bibinfo  {journal} {Proceedings of the Royal Society A}\ }\textbf {\bibinfo {volume} {470}},\ \bibinfo {pages} {20140394} (\bibinfo {year} {2014})}\BibitemShut {NoStop}%
\bibitem [{\citenamefont {Rybakov}\ and\ \citenamefont {Kiselev}(2019)}]{PhysRevB.99.064437}%
  \BibitemOpen
  \bibfield  {author} {\bibinfo {author} {\bibfnamefont {F.~N.}\ \bibnamefont {Rybakov}}\ and\ \bibinfo {author} {\bibfnamefont {N.~S.}\ \bibnamefont {Kiselev}},\ }\bibfield  {title} {\bibinfo {title} {Chiral magnetic skyrmions with arbitrary topological charge},\ }\href {https://doi.org/10.1103/PhysRevB.99.064437} {\bibfield  {journal} {\bibinfo  {journal} {Phys. Rev. B}\ }\textbf {\bibinfo {volume} {99}},\ \bibinfo {pages} {064437} (\bibinfo {year} {2019})}\BibitemShut {NoStop}%
\bibitem [{\citenamefont {Kuchkin}\ \emph {et~al.}(2020)\citenamefont {Kuchkin}, \citenamefont {Barton-Singer}, \citenamefont {Rybakov}, \citenamefont {Bl\"ugel}, \citenamefont {Schroers},\ and\ \citenamefont {Kiselev}}]{PhysRevB.102.144422}%
  \BibitemOpen
  \bibfield  {author} {\bibinfo {author} {\bibfnamefont {V.~M.}\ \bibnamefont {Kuchkin}}, \bibinfo {author} {\bibfnamefont {B.}~\bibnamefont {Barton-Singer}}, \bibinfo {author} {\bibfnamefont {F.~N.}\ \bibnamefont {Rybakov}}, \bibinfo {author} {\bibfnamefont {S.}~\bibnamefont {Bl\"ugel}}, \bibinfo {author} {\bibfnamefont {B.~J.}\ \bibnamefont {Schroers}},\ and\ \bibinfo {author} {\bibfnamefont {N.~S.}\ \bibnamefont {Kiselev}},\ }\bibfield  {title} {\bibinfo {title} {Magnetic skyrmions, chiral kinks, and holomorphic functions},\ }\href {https://doi.org/10.1103/PhysRevB.102.144422} {\bibfield  {journal} {\bibinfo  {journal} {Phys. Rev. B}\ }\textbf {\bibinfo {volume} {102}},\ \bibinfo {pages} {144422} (\bibinfo {year} {2020})}\BibitemShut {NoStop}%
\bibitem [{\citenamefont {Bernand-Mantel}\ \emph {et~al.}(2021)\citenamefont {Bernand-Mantel}, \citenamefont {Muratov},\ and\ \citenamefont {Simon}}]{Bernand-Mantel2021}%
  \BibitemOpen
  \bibfield  {author} {\bibinfo {author} {\bibfnamefont {A.}~\bibnamefont {Bernand-Mantel}}, \bibinfo {author} {\bibfnamefont {C.~B.}\ \bibnamefont {Muratov}},\ and\ \bibinfo {author} {\bibfnamefont {T.~M.}\ \bibnamefont {Simon}},\ }\bibfield  {title} {\bibinfo {title} {A quantitative description of skyrmions in ultrathin ferromagnetic films and rigidity of degree $\pm\,1$ harmonic maps from $\mathbb{R}^2$ to $\mathbb{S}^2$},\ }\href {https://doi.org/10.1007/s00205-020-01575-7} {\bibfield  {journal} {\bibinfo  {journal} {Archive for Rational Mechanics and Analysis}\ }\textbf {\bibinfo {volume} {239}},\ \bibinfo {pages} {219} (\bibinfo {year} {2021})}\BibitemShut {NoStop}%
\bibitem [{\citenamefont {Kirakosyan}\ and\ \citenamefont {Pokrovsky}(2006)}]{KIRAKOSYAN2006413}%
  \BibitemOpen
  \bibfield  {author} {\bibinfo {author} {\bibfnamefont {A.}~\bibnamefont {Kirakosyan}}\ and\ \bibinfo {author} {\bibfnamefont {V.}~\bibnamefont {Pokrovsky}},\ }\bibfield  {title} {\bibinfo {title} {From bubble to skyrmion: Dynamic transformation mediated by a strong magnetic tip},\ }\href {https://doi.org/10.1016/j.jmmm.2006.01.113} {\bibfield  {journal} {\bibinfo  {journal} {Journal of Magnetism and Magnetic Materials}\ }\textbf {\bibinfo {volume} {305}},\ \bibinfo {pages} {413} (\bibinfo {year} {2006})}\BibitemShut {NoStop}%
\bibitem [{\citenamefont {Leonov}\ and\ \citenamefont {Mostovoy}(2015)}]{Leonov2015}%
  \BibitemOpen
  \bibfield  {author} {\bibinfo {author} {\bibfnamefont {A.~O.}\ \bibnamefont {Leonov}}\ and\ \bibinfo {author} {\bibfnamefont {M.}~\bibnamefont {Mostovoy}},\ }\bibfield  {title} {\bibinfo {title} {Multiply periodic states and isolated skyrmions in an anisotropic frustrated magnet},\ }\href {https://doi.org/10.1038/ncomms9275} {\bibfield  {journal} {\bibinfo  {journal} {Nature Communications}\ }\textbf {\bibinfo {volume} {6}},\ \bibinfo {pages} {8275} (\bibinfo {year} {2015})}\BibitemShut {NoStop}%
\bibitem [{\citenamefont {R\'ozsa}\ \emph {et~al.}(2017)\citenamefont {R\'ozsa}, \citenamefont {Palot\'as}, \citenamefont {De\'ak}, \citenamefont {Simon}, \citenamefont {Yanes}, \citenamefont {Udvardi}, \citenamefont {Szunyogh},\ and\ \citenamefont {Nowak}}]{PhysRevB.95.094423}%
  \BibitemOpen
  \bibfield  {author} {\bibinfo {author} {\bibfnamefont {L.}~\bibnamefont {R\'ozsa}}, \bibinfo {author} {\bibfnamefont {K.}~\bibnamefont {Palot\'as}}, \bibinfo {author} {\bibfnamefont {A.}~\bibnamefont {De\'ak}}, \bibinfo {author} {\bibfnamefont {E.}~\bibnamefont {Simon}}, \bibinfo {author} {\bibfnamefont {R.}~\bibnamefont {Yanes}}, \bibinfo {author} {\bibfnamefont {L.}~\bibnamefont {Udvardi}}, \bibinfo {author} {\bibfnamefont {L.}~\bibnamefont {Szunyogh}},\ and\ \bibinfo {author} {\bibfnamefont {U.}~\bibnamefont {Nowak}},\ }\bibfield  {title} {\bibinfo {title} {Formation and stability of metastable skyrmionic spin structures with various topologies in an ultrathin film},\ }\href {https://doi.org/10.1103/PhysRevB.95.094423} {\bibfield  {journal} {\bibinfo  {journal} {Phys. Rev. B}\ }\textbf {\bibinfo {volume} {95}},\ \bibinfo {pages} {094423} (\bibinfo {year} {2017})}\BibitemShut {NoStop}%
\bibitem [{\citenamefont {Bogolubskaya}\ and\ \citenamefont {Bogolubsky}(1990)}]{Bogolubskaya1990}%
  \BibitemOpen
  \bibfield  {author} {\bibinfo {author} {\bibfnamefont {A.~A.}\ \bibnamefont {Bogolubskaya}}\ and\ \bibinfo {author} {\bibfnamefont {I.~L.}\ \bibnamefont {Bogolubsky}},\ }\bibfield  {title} {\bibinfo {title} {On stationary topological solitons in a two-dimensional anisotropic heisenberg model},\ }\href {https://doi.org/10.1007/BF01045888} {\bibfield  {journal} {\bibinfo  {journal} {Letters in Mathematical Physics}\ }\textbf {\bibinfo {volume} {19}},\ \bibinfo {pages} {171} (\bibinfo {year} {1990})}\BibitemShut {NoStop}%
\bibitem [{\citenamefont {Piette}\ \emph {et~al.}(1995)\citenamefont {Piette}, \citenamefont {Schroers},\ and\ \citenamefont {Zakrzewski}}]{Piette1995}%
  \BibitemOpen
  \bibfield  {author} {\bibinfo {author} {\bibfnamefont {B.~M. A.~G.}\ \bibnamefont {Piette}}, \bibinfo {author} {\bibfnamefont {B.~J.}\ \bibnamefont {Schroers}},\ and\ \bibinfo {author} {\bibfnamefont {W.~J.}\ \bibnamefont {Zakrzewski}},\ }\bibfield  {title} {\bibinfo {title} {Multisolitons in a two-dimensional skyrme model},\ }\href {https://doi.org/10.1007/BF01571317} {\bibfield  {journal} {\bibinfo  {journal} {Zeitschrift f{\"u}r Physik C Particles and Fields}\ }\textbf {\bibinfo {volume} {65}},\ \bibinfo {pages} {165} (\bibinfo {year} {1995})}\BibitemShut {NoStop}%
\bibitem [{\citenamefont {Ward}(1995)}]{Ward1995}%
  \BibitemOpen
  \bibfield  {author} {\bibinfo {author} {\bibfnamefont {R.~S.}\ \bibnamefont {Ward}},\ }\bibfield  {title} {\bibinfo {title} {Stable topological skyrmions on the {2D} lattice},\ }\href {https://doi.org/10.1007/BF00750845} {\bibfield  {journal} {\bibinfo  {journal} {Letters in Mathematical Physics}\ }\textbf {\bibinfo {volume} {35}},\ \bibinfo {pages} {385} (\bibinfo {year} {1995})}\BibitemShut {NoStop}%
\bibitem [{\citenamefont {Manton}\ and\ \citenamefont {Sutcliffe}(2004)}]{Manton_Sutcliffe_2004}%
  \BibitemOpen
  \bibfield  {author} {\bibinfo {author} {\bibfnamefont {N.}~\bibnamefont {Manton}}\ and\ \bibinfo {author} {\bibfnamefont {P.}~\bibnamefont {Sutcliffe}},\ }\href@noop {} {\emph {\bibinfo {title} {Topological Solitons}}}\ (\bibinfo  {publisher} {Cambridge University Press, New York},\ \bibinfo {year} {2004})\BibitemShut {NoStop}%
\bibitem [{\citenamefont {Hu}(1959)}]{HuHomotopyTheory}%
  \BibitemOpen
  \bibfield  {author} {\bibinfo {author} {\bibfnamefont {S.-T.}\ \bibnamefont {Hu}},\ }\href@noop {} {\emph {\bibinfo {title} {Homotopy theory}}}\ (\bibinfo  {publisher} {Academic Press, New York},\ \bibinfo {year} {1959})\BibitemShut {NoStop}%
\bibitem [{\citenamefont {Outerelo}\ and\ \citenamefont {Ruiz}(2009)}]{Mapping_degree_theory}%
  \BibitemOpen
  \bibfield  {author} {\bibinfo {author} {\bibfnamefont {E.}~\bibnamefont {Outerelo}}\ and\ \bibinfo {author} {\bibfnamefont {J.~M.}\ \bibnamefont {Ruiz}},\ }\href {https://doi.org/10.1090/gsm/108} {\emph {\bibinfo {title} {Mapping degree theory}}},\ Graduate Studies in Mathematics, Vol. 108\ (\bibinfo  {publisher} {American Mathematical Society Providence, Rhode Island},\ \bibinfo {year} {2009})\BibitemShut {NoStop}%
\bibitem [{\citenamefont {Skomski}(2008)}]{Skomski}%
  \BibitemOpen
  \bibfield  {author} {\bibinfo {author} {\bibfnamefont {R.}~\bibnamefont {Skomski}},\ }\href {https://doi.org/10.1093/acprof:oso/9780198570752.001.0001} {\emph {\bibinfo {title} {Simple Models of Magnetism}}}\ (\bibinfo  {publisher} {Oxford University Press},\ \bibinfo {year} {2008})\BibitemShut {NoStop}%
\bibitem [{\citenamefont {Gioia}\ and\ \citenamefont {James}(1997)}]{GioiaJames1997}%
  \BibitemOpen
  \bibfield  {author} {\bibinfo {author} {\bibfnamefont {G.}~\bibnamefont {Gioia}}\ and\ \bibinfo {author} {\bibfnamefont {R.~D.}\ \bibnamefont {James}},\ }\bibfield  {title} {\bibinfo {title} {Micromagnetics of very thin films},\ }\href {https://doi.org/10.1098/rspa.1997.0013} {\bibfield  {journal} {\bibinfo  {journal} {Proceedings of the Royal Society of London. Series A}\ }\textbf {\bibinfo {volume} {453}},\ \bibinfo {pages} {213} (\bibinfo {year} {1997})}\BibitemShut {NoStop}%
\bibitem [{\citenamefont {Slastikov}(2010)}]{Slastikov_config_aniso}%
  \BibitemOpen
  \bibfield  {author} {\bibinfo {author} {\bibfnamefont {V.~V.}\ \bibnamefont {Slastikov}},\ }\bibfield  {title} {\bibinfo {title} {A note on configurational anisotropy},\ }\href {https://doi.org/10.1098/rspa.2010.0070} {\bibfield  {journal} {\bibinfo  {journal} {Proceedings of the Royal Society A}\ }\textbf {\bibinfo {volume} {466}},\ \bibinfo {pages} {3167} (\bibinfo {year} {2010})}\BibitemShut {NoStop}%
\bibitem [{\citenamefont {Di~Fratta}\ \emph {et~al.}(2024)\citenamefont {Di~Fratta}, \citenamefont {Muratov},\ and\ \citenamefont {Slastikov}}]{ReducedEnergies2024}%
  \BibitemOpen
  \bibfield  {author} {\bibinfo {author} {\bibfnamefont {G.}~\bibnamefont {Di~Fratta}}, \bibinfo {author} {\bibfnamefont {C.~B.}\ \bibnamefont {Muratov}},\ and\ \bibinfo {author} {\bibfnamefont {V.~V.}\ \bibnamefont {Slastikov}},\ }\bibfield  {title} {\bibinfo {title} {Reduced energies for thin ferromagnetic films with perpendicular anisotropy},\ }\href {https://doi.org/10.1142/S0218202524500386} {\bibfield  {journal} {\bibinfo  {journal} {Mathematical Models and Methods in Applied Sciences}\ }\textbf {\bibinfo {volume} {34}},\ \bibinfo {pages} {1861} (\bibinfo {year} {2024})}\BibitemShut {NoStop}%
\bibitem [{\citenamefont {Volovik}\ and\ \citenamefont {Mineev}(1976)}]{Volovik_Mineev_1976}%
  \BibitemOpen
  \bibfield  {author} {\bibinfo {author} {\bibfnamefont {G.~E.}\ \bibnamefont {Volovik}}\ and\ \bibinfo {author} {\bibfnamefont {V.~P.}\ \bibnamefont {Mineev}},\ }\bibfield  {title} {\bibinfo {title} {Line and point singularities in superfluid $^3${He}},\ }\href {http://jetpletters.ru/ps/1818/article_27785.shtml} {\bibfield  {journal} {\bibinfo  {journal} {JETP Letters}\ }\textbf {\bibinfo {volume} {24}},\ \bibinfo {pages} {605} (\bibinfo {year} {1976})}\BibitemShut {NoStop}%
\bibitem [{\citenamefont {Monastyrsky}(1993)}]{Monastyrsky}%
  \BibitemOpen
  \bibfield  {author} {\bibinfo {author} {\bibfnamefont {M.}~\bibnamefont {Monastyrsky}},\ }\href {https://doi.org/10.1007/978-1-4899-2403-2} {\emph {\bibinfo {title} {Topology of Gauge Fields and Condensed Matter}}}\ (\bibinfo  {publisher} {Springer US},\ \bibinfo {year} {1993})\BibitemShut {NoStop}%
\bibitem [{\citenamefont {Volovik}\ and\ \citenamefont {Zhang}(2020)}]{PhysRevResearch.2.023263}%
  \BibitemOpen
  \bibfield  {author} {\bibinfo {author} {\bibfnamefont {G.~E.}\ \bibnamefont {Volovik}}\ and\ \bibinfo {author} {\bibfnamefont {K.}~\bibnamefont {Zhang}},\ }\bibfield  {title} {\bibinfo {title} {String monopoles, string walls, vortex skyrmions, and nexus objects in the polar distorted {$B$} phase of $^{3}\mathrm{He}$},\ }\href {https://doi.org/10.1103/PhysRevResearch.2.023263} {\bibfield  {journal} {\bibinfo  {journal} {Phys. Rev. Res.}\ }\textbf {\bibinfo {volume} {2}},\ \bibinfo {pages} {023263} (\bibinfo {year} {2020})}\BibitemShut {NoStop}%
\bibitem [{\citenamefont {Babaev}(2002)}]{PhysRevLett.89.067001}%
  \BibitemOpen
  \bibfield  {author} {\bibinfo {author} {\bibfnamefont {E.}~\bibnamefont {Babaev}},\ }\bibfield  {title} {\bibinfo {title} {Vortices with fractional flux in two-gap superconductors and in extended {Faddeev} model},\ }\href {https://doi.org/10.1103/PhysRevLett.89.067001} {\bibfield  {journal} {\bibinfo  {journal} {Phys. Rev. Lett.}\ }\textbf {\bibinfo {volume} {89}},\ \bibinfo {pages} {067001} (\bibinfo {year} {2002})}\BibitemShut {NoStop}%
\bibitem [{\citenamefont {Garaud}\ \emph {et~al.}(2013)\citenamefont {Garaud}, \citenamefont {Carlstr\"om}, \citenamefont {Babaev},\ and\ \citenamefont {Speight}}]{PhysRevB.87.014507}%
  \BibitemOpen
  \bibfield  {author} {\bibinfo {author} {\bibfnamefont {J.}~\bibnamefont {Garaud}}, \bibinfo {author} {\bibfnamefont {J.}~\bibnamefont {Carlstr\"om}}, \bibinfo {author} {\bibfnamefont {E.}~\bibnamefont {Babaev}},\ and\ \bibinfo {author} {\bibfnamefont {M.}~\bibnamefont {Speight}},\ }\bibfield  {title} {\bibinfo {title} {Chiral $\mathbb{C}{P}^{2}$ skyrmions in three-band superconductors},\ }\href {https://doi.org/10.1103/PhysRevB.87.014507} {\bibfield  {journal} {\bibinfo  {journal} {Phys. Rev. B}\ }\textbf {\bibinfo {volume} {87}},\ \bibinfo {pages} {014507} (\bibinfo {year} {2013})}\BibitemShut {NoStop}%
\bibitem [{\citenamefont {Belavin}\ and\ \citenamefont {Polyakov}(1975)}]{BP1975}%
  \BibitemOpen
  \bibfield  {author} {\bibinfo {author} {\bibfnamefont {A.~A.}\ \bibnamefont {Belavin}}\ and\ \bibinfo {author} {\bibfnamefont {A.~M.}\ \bibnamefont {Polyakov}},\ }\bibfield  {title} {\bibinfo {title} {Metastable states of two-dimensional isotropic ferromagnets},\ }\href {http://jetpletters.ru/ps/1529/article_23383.shtml} {\bibfield  {journal} {\bibinfo  {journal} {JETP Lett.}\ }\textbf {\bibinfo {volume} {22}},\ \bibinfo {pages} {245} (\bibinfo {year} {1975})}\BibitemShut {NoStop}%
\bibitem [{\citenamefont {Mineev}(1998)}]{Mineev_book}%
  \BibitemOpen
  \bibfield  {author} {\bibinfo {author} {\bibfnamefont {V.~P.}\ \bibnamefont {Mineev}},\ }\href@noop {} {\emph {\bibinfo {title} {Topologically stable defects and solitons in ordered media}}}\ (\bibinfo  {publisher} {Harwood Academic Publishers, Amsterdam},\ \bibinfo {year} {1998})\BibitemShut {NoStop}%
\bibitem [{\citenamefont {Kleman}\ and\ \citenamefont {Lavrentovich}(2003)}]{Kleman_Lavrentovich_2003}%
  \BibitemOpen
  \bibinfo {editor} {\bibfnamefont {M.}~\bibnamefont {Kleman}}\ and\ \bibinfo {editor} {\bibfnamefont {O.~D.}\ \bibnamefont {Lavrentovich}},\ eds.,\ \href {https://doi.org/10.1007/b97416} {\emph {\bibinfo {title} {Soft Matter Physics: an introduction}}},\ Partially Ordered Systems\ (\bibinfo  {publisher} {Springer, New York},\ \bibinfo {year} {2003})\BibitemShut {NoStop}%
\bibitem [{\citenamefont {Kléman}\ \emph {et~al.}(1977)\citenamefont {Kléman}, \citenamefont {Michel},\ and\ \citenamefont {Toulouse}}]{KlemanMichelToulouse_1977}%
  \BibitemOpen
  \bibfield  {author} {\bibinfo {author} {\bibfnamefont {M.}~\bibnamefont {Kléman}}, \bibinfo {author} {\bibfnamefont {L.}~\bibnamefont {Michel}},\ and\ \bibinfo {author} {\bibfnamefont {G.}~\bibnamefont {Toulouse}},\ }\bibfield  {title} {\bibinfo {title} {Classification of topologically stable defects in ordered media},\ }\href {https://doi.org/10.1051/jphyslet:019770038010019500} {\bibfield  {journal} {\bibinfo  {journal} {J. Physique Lett.}\ }\textbf {\bibinfo {volume} {38}},\ \bibinfo {pages} {195} (\bibinfo {year} {1977})}\BibitemShut {NoStop}%
\bibitem [{\citenamefont {Mermin}(1979)}]{RevModPhys.51.591}%
  \BibitemOpen
  \bibfield  {author} {\bibinfo {author} {\bibfnamefont {N.~D.}\ \bibnamefont {Mermin}},\ }\bibfield  {title} {\bibinfo {title} {The topological theory of defects in ordered media},\ }\href {https://doi.org/10.1103/RevModPhys.51.591} {\bibfield  {journal} {\bibinfo  {journal} {Rev. Mod. Phys.}\ }\textbf {\bibinfo {volume} {51}},\ \bibinfo {pages} {591} (\bibinfo {year} {1979})}\BibitemShut {NoStop}%
\bibitem [{\citenamefont {Wu}\ and\ \citenamefont {Smalyukh}(2022)}]{Wu_Smalyukh_review_2022}%
  \BibitemOpen
  \bibfield  {author} {\bibinfo {author} {\bibfnamefont {J.-S.}\ \bibnamefont {Wu}}\ and\ \bibinfo {author} {\bibfnamefont {I.~I.}\ \bibnamefont {Smalyukh}},\ }\bibfield  {title} {\bibinfo {title} {Hopfions, heliknotons, skyrmions, torons and both abelian and nonabelian vortices in chiral liquid crystals},\ }\href {https://doi.org/10.1080/21680396.2022.2040058} {\bibfield  {journal} {\bibinfo  {journal} {Liquid Crystals Reviews}\ }\textbf {\bibinfo {volume} {10}},\ \bibinfo {pages} {34} (\bibinfo {year} {2022})}\BibitemShut {NoStop}%
\bibitem [{\citenamefont {Rybakov}\ and\ \citenamefont {Eriksson}(2022)}]{Rybakov_Eriksson}%
  \BibitemOpen
  \bibfield  {author} {\bibinfo {author} {\bibfnamefont {F.~N.}\ \bibnamefont {Rybakov}}\ and\ \bibinfo {author} {\bibfnamefont {O.}~\bibnamefont {Eriksson}},\ }\bibfield  {title} {\bibinfo {title} {Non-abelian vortices in magnets},\ }\href {https://doi.org/10.48550/arXiv.2205.15264} {\bibfield  {journal} {\bibinfo  {journal} {arXiv:2205.15264}\ } (\bibinfo {year} {2022})}\BibitemShut {NoStop}%
\bibitem [{\citenamefont {Halcrow}\ and\ \citenamefont {Babaev}(2024)}]{PhysRevResearch.6.L032011}%
  \BibitemOpen
  \bibfield  {author} {\bibinfo {author} {\bibfnamefont {C.}~\bibnamefont {Halcrow}}\ and\ \bibinfo {author} {\bibfnamefont {E.}~\bibnamefont {Babaev}},\ }\bibfield  {title} {\bibinfo {title} {Fractional skyrme lines in ferroelectric barium titanate},\ }\href {https://doi.org/10.1103/PhysRevResearch.6.L032011} {\bibfield  {journal} {\bibinfo  {journal} {Phys. Rev. Res.}\ }\textbf {\bibinfo {volume} {6}},\ \bibinfo {pages} {L032011} (\bibinfo {year} {2024})}\BibitemShut {NoStop}%
\bibitem [{\citenamefont {del Ser}\ \emph {et~al.}(2024)\citenamefont {del Ser}, \citenamefont {El~Achchi},\ and\ \citenamefont {Rosch}}]{PhysRevB.110.094442}%
  \BibitemOpen
  \bibfield  {author} {\bibinfo {author} {\bibfnamefont {N.}~\bibnamefont {del Ser}}, \bibinfo {author} {\bibfnamefont {I.}~\bibnamefont {El~Achchi}},\ and\ \bibinfo {author} {\bibfnamefont {A.}~\bibnamefont {Rosch}},\ }\bibfield  {title} {\bibinfo {title} {Fractional topological charges in two-dimensional magnets},\ }\href {https://doi.org/10.1103/PhysRevB.110.094442} {\bibfield  {journal} {\bibinfo  {journal} {Phys. Rev. B}\ }\textbf {\bibinfo {volume} {110}},\ \bibinfo {pages} {094442} (\bibinfo {year} {2024})}\BibitemShut {NoStop}%
\bibitem [{\citenamefont {Guslienko}\ \emph {et~al.}(2008)\citenamefont {Guslienko}, \citenamefont {Lee},\ and\ \citenamefont {Kim}}]{PhysRevLett.100.027203}%
  \BibitemOpen
  \bibfield  {author} {\bibinfo {author} {\bibfnamefont {K.~Y.}\ \bibnamefont {Guslienko}}, \bibinfo {author} {\bibfnamefont {K.-S.}\ \bibnamefont {Lee}},\ and\ \bibinfo {author} {\bibfnamefont {S.-K.}\ \bibnamefont {Kim}},\ }\bibfield  {title} {\bibinfo {title} {Dynamic origin of vortex core switching in soft magnetic nanodots},\ }\href {https://doi.org/10.1103/PhysRevLett.100.027203} {\bibfield  {journal} {\bibinfo  {journal} {Phys. Rev. Lett.}\ }\textbf {\bibinfo {volume} {100}},\ \bibinfo {pages} {027203} (\bibinfo {year} {2008})}\BibitemShut {NoStop}%
\bibitem [{\citenamefont {Moutafis}\ \emph {et~al.}(2009)\citenamefont {Moutafis}, \citenamefont {Komineas},\ and\ \citenamefont {Bland}}]{PhysRevB.79.224429}%
  \BibitemOpen
  \bibfield  {author} {\bibinfo {author} {\bibfnamefont {C.}~\bibnamefont {Moutafis}}, \bibinfo {author} {\bibfnamefont {S.}~\bibnamefont {Komineas}},\ and\ \bibinfo {author} {\bibfnamefont {J.~A.~C.}\ \bibnamefont {Bland}},\ }\bibfield  {title} {\bibinfo {title} {Dynamics and switching processes for magnetic bubbles in nanoelements},\ }\href {https://doi.org/10.1103/PhysRevB.79.224429} {\bibfield  {journal} {\bibinfo  {journal} {Phys. Rev. B}\ }\textbf {\bibinfo {volume} {79}},\ \bibinfo {pages} {224429} (\bibinfo {year} {2009})}\BibitemShut {NoStop}%
\bibitem [{\citenamefont {Heo}\ \emph {et~al.}(2016)\citenamefont {Heo}, \citenamefont {Kiselev}, \citenamefont {Nandy}, \citenamefont {Bl{\"u}gel},\ and\ \citenamefont {Rasing}}]{heo2016switching}%
  \BibitemOpen
  \bibfield  {author} {\bibinfo {author} {\bibfnamefont {C.}~\bibnamefont {Heo}}, \bibinfo {author} {\bibfnamefont {N.~S.}\ \bibnamefont {Kiselev}}, \bibinfo {author} {\bibfnamefont {A.~K.}\ \bibnamefont {Nandy}}, \bibinfo {author} {\bibfnamefont {S.}~\bibnamefont {Bl{\"u}gel}},\ and\ \bibinfo {author} {\bibfnamefont {T.}~\bibnamefont {Rasing}},\ }\bibfield  {title} {\bibinfo {title} {Switching of chiral magnetic skyrmions by picosecond magnetic field pulses via transient topological states},\ }\href {https://doi.org/10.1038/srep27146} {\bibfield  {journal} {\bibinfo  {journal} {Scientific reports}\ }\textbf {\bibinfo {volume} {6}},\ \bibinfo {pages} {27146} (\bibinfo {year} {2016})}\BibitemShut {NoStop}%
\bibitem [{\citenamefont {Bessarab}\ \emph {et~al.}(2018)\citenamefont {Bessarab}, \citenamefont {M{\"u}ller}, \citenamefont {Lobanov}, \citenamefont {Rybakov}, \citenamefont {Kiselev}, \citenamefont {J{\'o}nsson}, \citenamefont {Uzdin}, \citenamefont {Bl{\"u}gel}, \citenamefont {Bergqvist},\ and\ \citenamefont {Delin}}]{Bessarab2018}%
  \BibitemOpen
  \bibfield  {author} {\bibinfo {author} {\bibfnamefont {P.~F.}\ \bibnamefont {Bessarab}}, \bibinfo {author} {\bibfnamefont {G.~P.}\ \bibnamefont {M{\"u}ller}}, \bibinfo {author} {\bibfnamefont {I.~S.}\ \bibnamefont {Lobanov}}, \bibinfo {author} {\bibfnamefont {F.~N.}\ \bibnamefont {Rybakov}}, \bibinfo {author} {\bibfnamefont {N.~S.}\ \bibnamefont {Kiselev}}, \bibinfo {author} {\bibfnamefont {H.}~\bibnamefont {J{\'o}nsson}}, \bibinfo {author} {\bibfnamefont {V.~M.}\ \bibnamefont {Uzdin}}, \bibinfo {author} {\bibfnamefont {S.}~\bibnamefont {Bl{\"u}gel}}, \bibinfo {author} {\bibfnamefont {L.}~\bibnamefont {Bergqvist}},\ and\ \bibinfo {author} {\bibfnamefont {A.}~\bibnamefont {Delin}},\ }\bibfield  {title} {\bibinfo {title} {Lifetime of racetrack skyrmions},\ }\href {https://doi.org/10.1038/s41598-018-21623-3} {\bibfield  {journal} {\bibinfo  {journal} {Scientific Reports}\ }\textbf {\bibinfo {volume} {8}},\ \bibinfo {pages} {3433} (\bibinfo {year} {2018})}\BibitemShut {NoStop}%
\bibitem [{\citenamefont {Muckel}\ \emph {et~al.}(2021)\citenamefont {Muckel}, \citenamefont {von Malottki}, \citenamefont {Holl}, \citenamefont {Pestka}, \citenamefont {Pratzer}, \citenamefont {Bessarab}, \citenamefont {Heinze},\ and\ \citenamefont {Morgenstern}}]{Muckel2021}%
  \BibitemOpen
  \bibfield  {author} {\bibinfo {author} {\bibfnamefont {F.}~\bibnamefont {Muckel}}, \bibinfo {author} {\bibfnamefont {S.}~\bibnamefont {von Malottki}}, \bibinfo {author} {\bibfnamefont {C.}~\bibnamefont {Holl}}, \bibinfo {author} {\bibfnamefont {B.}~\bibnamefont {Pestka}}, \bibinfo {author} {\bibfnamefont {M.}~\bibnamefont {Pratzer}}, \bibinfo {author} {\bibfnamefont {P.~F.}\ \bibnamefont {Bessarab}}, \bibinfo {author} {\bibfnamefont {S.}~\bibnamefont {Heinze}},\ and\ \bibinfo {author} {\bibfnamefont {M.}~\bibnamefont {Morgenstern}},\ }\bibfield  {title} {\bibinfo {title} {Experimental identification of two distinct skyrmion collapse mechanisms},\ }\href {https://doi.org/10.1038/s41567-020-01101-2} {\bibfield  {journal} {\bibinfo  {journal} {Nature Physics}\ }\textbf {\bibinfo {volume} {17}},\ \bibinfo {pages} {395} (\bibinfo {year} {2021})}\BibitemShut {NoStop}%
\bibitem [{\citenamefont {Bernand-Mantel}\ \emph {et~al.}(2022)\citenamefont {Bernand-Mantel}, \citenamefont {Muratov},\ and\ \citenamefont {Slastikov}}]{MicromagneticLifetime2022}%
  \BibitemOpen
  \bibfield  {author} {\bibinfo {author} {\bibfnamefont {A.}~\bibnamefont {Bernand-Mantel}}, \bibinfo {author} {\bibfnamefont {C.~B.}\ \bibnamefont {Muratov}},\ and\ \bibinfo {author} {\bibfnamefont {V.~V.}\ \bibnamefont {Slastikov}},\ }\bibfield  {title} {\bibinfo {title} {A micromagnetic theory of skyrmion lifetime in ultrathin ferromagnetic films},\ }\href {https://doi.org/10.1073/pnas.2122237119} {\bibfield  {journal} {\bibinfo  {journal} {Proceedings of the National Academy of Sciences}\ }\textbf {\bibinfo {volume} {119}},\ \bibinfo {pages} {e2122237119} (\bibinfo {year} {2022})}\BibitemShut {NoStop}%
\bibitem [{\citenamefont {Dubrovin}\ \emph {et~al.}(1985)\citenamefont {Dubrovin}, \citenamefont {Fomenko},\ and\ \citenamefont {Novikov}}]{DubrovinFomenkoNovikov_2}%
  \BibitemOpen
  \bibfield  {author} {\bibinfo {author} {\bibfnamefont {B.~A.}\ \bibnamefont {Dubrovin}}, \bibinfo {author} {\bibfnamefont {A.~T.}\ \bibnamefont {Fomenko}},\ and\ \bibinfo {author} {\bibfnamefont {S.~P.}\ \bibnamefont {Novikov}},\ }\href {https://doi.org/10.1007/978-1-4612-1100-6} {\emph {\bibinfo {title} {Modern Geometry -- Methods and Applications. Part II. The Geometry and Topology of Manifolds}}},\ Graduate Texts in Mathematics, Vol. 104\ (\bibinfo  {publisher} {Springer-Verlag, New York},\ \bibinfo {year} {1985})\BibitemShut {NoStop}%
\bibitem [{\citenamefont {Kronecker}(1869)}]{Kronecker_1869}%
  \BibitemOpen
  \bibfield  {author} {\bibinfo {author} {\bibfnamefont {L.}~\bibnamefont {Kronecker}},\ }\bibfield  {title} {\bibinfo {title} {Über systeme von functionen mehrer variabeln},\ }\href@noop {} {\bibfield  {journal} {\bibinfo  {journal} {Monatsberichte königlich Preuss. Akad. Wissens. Berlin}\ } (\bibinfo {year} {1869})}\BibitemShut {NoStop}%
\bibitem [{\citenamefont {Flanders}(1963)}]{Flanders_DifferentialForms}%
  \BibitemOpen
  \bibfield  {author} {\bibinfo {author} {\bibfnamefont {H.}~\bibnamefont {Flanders}},\ }\href@noop {} {\emph {\bibinfo {title} {Differential Forms with Applications to the Physical Sciences}}}\ (\bibinfo  {publisher} {Academic Press},\ \bibinfo {year} {1963})\BibitemShut {NoStop}%
\bibitem [{\citenamefont {Strom}(2011)}]{ModernClassicalHomotopyTheory}%
  \BibitemOpen
  \bibfield  {author} {\bibinfo {author} {\bibfnamefont {J.}~\bibnamefont {Strom}},\ }\href@noop {} {\emph {\bibinfo {title} {Modern Classical Homotopy Theory}}},\ Graduate Studies in Mathematics, Vol. 127\ (\bibinfo  {publisher} {American Mathematical Society, Providence, RI},\ \bibinfo {year} {2011})\BibitemShut {NoStop}%
\bibitem [{\citenamefont {Hilton}(1955)}]{Hilton1955}%
  \BibitemOpen
  \bibfield  {author} {\bibinfo {author} {\bibfnamefont {P.~J.}\ \bibnamefont {Hilton}},\ }\bibfield  {title} {\bibinfo {title} {On the homotopy groups of the union of spheres},\ }\href {https://doi.org/10.1112/jlms/s1-30.2.154} {\bibfield  {journal} {\bibinfo  {journal} {Journal of the London Mathematical Society}\ }\textbf {\bibinfo {volume} {s1-30}},\ \bibinfo {pages} {154} (\bibinfo {year} {1955})}\BibitemShut {NoStop}%
\bibitem [{\citenamefont {Bott}\ and\ \citenamefont {Tu}(1982)}]{BottTu1982}%
  \BibitemOpen
  \bibfield  {author} {\bibinfo {author} {\bibfnamefont {R.}~\bibnamefont {Bott}}\ and\ \bibinfo {author} {\bibfnamefont {L.~W.}\ \bibnamefont {Tu}},\ }\href {https://doi.org/10.1007/978-1-4757-3951-0} {\emph {\bibinfo {title} {Differential Forms in Algebraic Topology}}},\ Graduate Texts in Mathematics, Vol. 82\ (\bibinfo  {publisher} {Springer New York, NY},\ \bibinfo {year} {1982})\BibitemShut {NoStop}%
\bibitem [{\citenamefont {Rybakov}(2021)}]{Rybakov_thesis}%
  \BibitemOpen
  \bibfield  {author} {\bibinfo {author} {\bibfnamefont {F.~N.}\ \bibnamefont {Rybakov}},\ }\emph {\bibinfo {title} {Topological excitations in field theory models of superconductivity and magnetism}},\ \href {http://urn.kb.se/resolve?urn=urn:nbn:se:kth:diva-301652} {Ph.D. thesis},\ \bibinfo  {school} {KTH Royal Institute of Technology}, \bibinfo {address} {Stockholm} (\bibinfo {year} {2021})\BibitemShut {NoStop}%
\bibitem [{\citenamefont {Zheng}\ \emph {et~al.}(2023)\citenamefont {Zheng}, \citenamefont {Kiselev}, \citenamefont {Rybakov}, \citenamefont {Yang}, \citenamefont {Shi}, \citenamefont {Bl{\"u}gel},\ and\ \citenamefont {Dunin-Borkowski}}]{Zheng2023}%
  \BibitemOpen
  \bibfield  {author} {\bibinfo {author} {\bibfnamefont {F.}~\bibnamefont {Zheng}}, \bibinfo {author} {\bibfnamefont {N.~S.}\ \bibnamefont {Kiselev}}, \bibinfo {author} {\bibfnamefont {F.~N.}\ \bibnamefont {Rybakov}}, \bibinfo {author} {\bibfnamefont {L.}~\bibnamefont {Yang}}, \bibinfo {author} {\bibfnamefont {W.}~\bibnamefont {Shi}}, \bibinfo {author} {\bibfnamefont {S.}~\bibnamefont {Bl{\"u}gel}},\ and\ \bibinfo {author} {\bibfnamefont {R.~E.}\ \bibnamefont {Dunin-Borkowski}},\ }\bibfield  {title} {\bibinfo {title} {Hopfion rings in a cubic chiral magnet},\ }\href {https://doi.org/10.1038/s41586-023-06658-5} {\bibfield  {journal} {\bibinfo  {journal} {Nature}\ }\textbf {\bibinfo {volume} {623}},\ \bibinfo {pages} {718} (\bibinfo {year} {2023})}\BibitemShut {NoStop}%
\bibitem [{\citenamefont {Barton-Singer}\ \emph {et~al.}(2020)\citenamefont {Barton-Singer}, \citenamefont {Ross},\ and\ \citenamefont {Schroers}}]{Barton-Singer2020}%
  \BibitemOpen
  \bibfield  {author} {\bibinfo {author} {\bibfnamefont {B.}~\bibnamefont {Barton-Singer}}, \bibinfo {author} {\bibfnamefont {C.}~\bibnamefont {Ross}},\ and\ \bibinfo {author} {\bibfnamefont {B.~J.}\ \bibnamefont {Schroers}},\ }\bibfield  {title} {\bibinfo {title} {Magnetic skyrmions at critical coupling},\ }\href {https://doi.org/10.1007/s00220-019-03676-1} {\bibfield  {journal} {\bibinfo  {journal} {Communications in Mathematical Physics}\ }\textbf {\bibinfo {volume} {375}},\ \bibinfo {pages} {2259} (\bibinfo {year} {2020})}\BibitemShut {NoStop}%
\bibitem [{\citenamefont {Ohara}\ \emph {et~al.}(2022)\citenamefont {Ohara}, \citenamefont {Zhang}, \citenamefont {Chen}, \citenamefont {Kato}, \citenamefont {Xia}, \citenamefont {Ezawa}, \citenamefont {Tretiakov}, \citenamefont {Hou}, \citenamefont {Zhou}, \citenamefont {Zhao}, \citenamefont {Yang},\ and\ \citenamefont {Liu}}]{Ohara2022}%
  \BibitemOpen
  \bibfield  {author} {\bibinfo {author} {\bibfnamefont {K.}~\bibnamefont {Ohara}}, \bibinfo {author} {\bibfnamefont {X.}~\bibnamefont {Zhang}}, \bibinfo {author} {\bibfnamefont {Y.}~\bibnamefont {Chen}}, \bibinfo {author} {\bibfnamefont {S.}~\bibnamefont {Kato}}, \bibinfo {author} {\bibfnamefont {J.}~\bibnamefont {Xia}}, \bibinfo {author} {\bibfnamefont {M.}~\bibnamefont {Ezawa}}, \bibinfo {author} {\bibfnamefont {O.~A.}\ \bibnamefont {Tretiakov}}, \bibinfo {author} {\bibfnamefont {Z.}~\bibnamefont {Hou}}, \bibinfo {author} {\bibfnamefont {Y.}~\bibnamefont {Zhou}}, \bibinfo {author} {\bibfnamefont {G.}~\bibnamefont {Zhao}}, \bibinfo {author} {\bibfnamefont {J.}~\bibnamefont {Yang}},\ and\ \bibinfo {author} {\bibfnamefont {X.}~\bibnamefont {Liu}},\ }\bibfield  {title} {\bibinfo {title} {Reversible transformation between isolated skyrmions and bimerons},\ }\href {https://doi.org/10.1021/acs.nanolett.2c03106} {\bibfield  {journal} {\bibinfo  {journal} {Nano Letters}\ }\textbf {\bibinfo {volume} {22}},\ \bibinfo
  {pages} {8559} (\bibinfo {year} {2022})}\BibitemShut {NoStop}%
\bibitem [{\citenamefont {Berg}\ and\ \citenamefont {Lüscher}(1981)}]{BERG1981412}%
  \BibitemOpen
  \bibfield  {author} {\bibinfo {author} {\bibfnamefont {B.}~\bibnamefont {Berg}}\ and\ \bibinfo {author} {\bibfnamefont {M.}~\bibnamefont {Lüscher}},\ }\bibfield  {title} {\bibinfo {title} {Definition and statistical distributions of a topological number in the lattice {$O(3)$} $\sigma$-model},\ }\href {https://doi.org/10.1016/0550-3213(81)90568-X} {\bibfield  {journal} {\bibinfo  {journal} {Nuclear Physics B}\ }\textbf {\bibinfo {volume} {190}},\ \bibinfo {pages} {412} (\bibinfo {year} {1981})}\BibitemShut {NoStop}%
\bibitem [{\citenamefont {Fulton}(1995)}]{Fulton_AlgebraicTopology}%
  \BibitemOpen
  \bibfield  {author} {\bibinfo {author} {\bibfnamefont {W.}~\bibnamefont {Fulton}},\ }\href {https://doi.org/10.1007/978-1-4612-4180-5} {\emph {\bibinfo {title} {Algebraic Topology: A First Course}}},\ Graduate Texts in Mathematics, Vol. 153\ (\bibinfo  {publisher} {Springer, New York, NY},\ \bibinfo {year} {1995})\BibitemShut {NoStop}%
\bibitem [{\citenamefont {Garanin}(1997)}]{PhysRevB.55.3050}%
  \BibitemOpen
  \bibfield  {author} {\bibinfo {author} {\bibfnamefont {D.~A.}\ \bibnamefont {Garanin}},\ }\bibfield  {title} {\bibinfo {title} {Fokker-planck and landau-lifshitz-bloch equations for classical ferromagnets},\ }\href {https://doi.org/10.1103/PhysRevB.55.3050} {\bibfield  {journal} {\bibinfo  {journal} {Phys. Rev. B}\ }\textbf {\bibinfo {volume} {55}},\ \bibinfo {pages} {3050} (\bibinfo {year} {1997})}\BibitemShut {NoStop}%
\bibitem [{\citenamefont {R{\"o}{\ss}ler}\ \emph {et~al.}(2006)\citenamefont {R{\"o}{\ss}ler}, \citenamefont {Bogdanov},\ and\ \citenamefont {Pfleiderer}}]{Ulrich_Alex_Christian_2006}%
  \BibitemOpen
  \bibfield  {author} {\bibinfo {author} {\bibfnamefont {U.~K.}\ \bibnamefont {R{\"o}{\ss}ler}}, \bibinfo {author} {\bibfnamefont {A.~N.}\ \bibnamefont {Bogdanov}},\ and\ \bibinfo {author} {\bibfnamefont {C.}~\bibnamefont {Pfleiderer}},\ }\bibfield  {title} {\bibinfo {title} {Spontaneous skyrmion ground states in magnetic metals},\ }\href {https://doi.org/10.1038/nature05056} {\bibfield  {journal} {\bibinfo  {journal} {Nature}\ }\textbf {\bibinfo {volume} {442}},\ \bibinfo {pages} {797} (\bibinfo {year} {2006})}\BibitemShut {NoStop}%
\bibitem [{\citenamefont {Aharoni}(2000)}]{Aharoni_book}%
  \BibitemOpen
  \bibfield  {author} {\bibinfo {author} {\bibfnamefont {A.}~\bibnamefont {Aharoni}},\ }\href {https://doi.org/10.1093/oso/9780198508083.001.0001} {\emph {\bibinfo {title} {Introduction to the Theory of Ferromagnetism}}},\ International Series of Monographs on Physics, Vol. 109\ (\bibinfo  {publisher} {Oxford University Press},\ \bibinfo {year} {2000})\BibitemShut {NoStop}%
\bibitem [{\citenamefont {Chubykalo-Fesenko}\ and\ \citenamefont {Nieves}(2020)}]{Chubykalo-Fesenko2020}%
  \BibitemOpen
  \bibfield  {author} {\bibinfo {author} {\bibfnamefont {O.}~\bibnamefont {Chubykalo-Fesenko}}\ and\ \bibinfo {author} {\bibfnamefont {P.}~\bibnamefont {Nieves}},\ }\bibinfo {title} {{Landau-Lifshitz-Bloch} approach for magnetization dynamics close to phase transition},\ in\ \href {https://doi.org/10.1007/978-3-319-44677-6_72} {\emph {\bibinfo {booktitle} {Handbook of Materials Modeling. Methods: Theory and Modeling}}},\ \bibinfo {editor} {edited by\ \bibinfo {editor} {\bibfnamefont {W.}~\bibnamefont {Andreoni}}\ and\ \bibinfo {editor} {\bibfnamefont {S.}~\bibnamefont {Yip}}}\ (\bibinfo  {publisher} {Springer International Publishing},\ \bibinfo {address} {Cham},\ \bibinfo {year} {2020})\ pp.\ \bibinfo {pages} {867--893}\BibitemShut {NoStop}%
\bibitem [{\citenamefont {Lines}\ and\ \citenamefont {Glass}(1977)}]{Lines_Glass_Ferroelectrics}%
  \BibitemOpen
  \bibfield  {author} {\bibinfo {author} {\bibfnamefont {M.~E.}\ \bibnamefont {Lines}}\ and\ \bibinfo {author} {\bibfnamefont {A.~M.}\ \bibnamefont {Glass}},\ }\href {https://doi.org/10.1093/acprof:oso/9780198507789.001.0001} {\emph {\bibinfo {title} {Principles and Applications of Ferroelectrics and Related Materials}}}\ (\bibinfo  {publisher} {Clarendon Press, Oxford},\ \bibinfo {year} {1977})\BibitemShut {NoStop}%
\bibitem [{\citenamefont {Hlinka}\ and\ \citenamefont {M\'arton}(2006)}]{PhysRevB.74.104104}%
  \BibitemOpen
  \bibfield  {author} {\bibinfo {author} {\bibfnamefont {J.}~\bibnamefont {Hlinka}}\ and\ \bibinfo {author} {\bibfnamefont {P.}~\bibnamefont {M\'arton}},\ }\bibfield  {title} {\bibinfo {title} {Phenomenological model of a 90$^\circ$ domain wall in $\mathrm{Ba}\mathrm{Ti}\mathrm{O}_{3}$-type ferroelectrics},\ }\href {https://doi.org/10.1103/PhysRevB.74.104104} {\bibfield  {journal} {\bibinfo  {journal} {Phys. Rev. B}\ }\textbf {\bibinfo {volume} {74}},\ \bibinfo {pages} {104104} (\bibinfo {year} {2006})}\BibitemShut {NoStop}%
\bibitem [{\citenamefont {Das}\ \emph {et~al.}(2019)\citenamefont {Das}, \citenamefont {Tang}, \citenamefont {Hong}, \citenamefont {Gon{\c{c}}alves}, \citenamefont {McCarter}, \citenamefont {Klewe}, \citenamefont {Nguyen}, \citenamefont {G{\'o}mez-Ortiz}, \citenamefont {Shafer}, \citenamefont {Arenholz}, \citenamefont {Stoica}, \citenamefont {Hsu}, \citenamefont {Wang}, \citenamefont {Ophus}, \citenamefont {Liu}, \citenamefont {Nelson}, \citenamefont {Saremi}, \citenamefont {Prasad}, \citenamefont {Mei}, \citenamefont {Schlom}, \citenamefont {{\'I}{\~{n}}iguez}, \citenamefont {Garc{\'i}a-Fern{\'a}ndez}, \citenamefont {Muller}, \citenamefont {Chen}, \citenamefont {Junquera}, \citenamefont {Martin},\ and\ \citenamefont {Ramesh}}]{Das2019}%
  \BibitemOpen
  \bibfield  {author} {\bibinfo {author} {\bibfnamefont {S.}~\bibnamefont {Das}}, \bibinfo {author} {\bibfnamefont {Y.~L.}\ \bibnamefont {Tang}}, \bibinfo {author} {\bibfnamefont {Z.}~\bibnamefont {Hong}}, \bibinfo {author} {\bibfnamefont {M.~A.~P.}\ \bibnamefont {Gon{\c{c}}alves}}, \bibinfo {author} {\bibfnamefont {M.~R.}\ \bibnamefont {McCarter}}, \bibinfo {author} {\bibfnamefont {C.}~\bibnamefont {Klewe}}, \bibinfo {author} {\bibfnamefont {K.~X.}\ \bibnamefont {Nguyen}}, \bibinfo {author} {\bibfnamefont {F.}~\bibnamefont {G{\'o}mez-Ortiz}}, \bibinfo {author} {\bibfnamefont {P.}~\bibnamefont {Shafer}}, \bibinfo {author} {\bibfnamefont {E.}~\bibnamefont {Arenholz}}, \bibinfo {author} {\bibfnamefont {V.~A.}\ \bibnamefont {Stoica}}, \bibinfo {author} {\bibfnamefont {S.-L.}\ \bibnamefont {Hsu}}, \bibinfo {author} {\bibfnamefont {B.}~\bibnamefont {Wang}}, \bibinfo {author} {\bibfnamefont {C.}~\bibnamefont {Ophus}}, \bibinfo {author} {\bibfnamefont {J.~F.}\ \bibnamefont {Liu}}, \bibinfo {author} {\bibfnamefont
  {C.~T.}\ \bibnamefont {Nelson}}, \bibinfo {author} {\bibfnamefont {S.}~\bibnamefont {Saremi}}, \bibinfo {author} {\bibfnamefont {B.}~\bibnamefont {Prasad}}, \bibinfo {author} {\bibfnamefont {A.~B.}\ \bibnamefont {Mei}}, \bibinfo {author} {\bibfnamefont {D.~G.}\ \bibnamefont {Schlom}}, \bibinfo {author} {\bibfnamefont {J.}~\bibnamefont {{\'I}{\~{n}}iguez}}, \bibinfo {author} {\bibfnamefont {P.}~\bibnamefont {Garc{\'i}a-Fern{\'a}ndez}}, \bibinfo {author} {\bibfnamefont {D.~A.}\ \bibnamefont {Muller}}, \bibinfo {author} {\bibfnamefont {L.~Q.}\ \bibnamefont {Chen}}, \bibinfo {author} {\bibfnamefont {J.}~\bibnamefont {Junquera}}, \bibinfo {author} {\bibfnamefont {L.~W.}\ \bibnamefont {Martin}},\ and\ \bibinfo {author} {\bibfnamefont {R.}~\bibnamefont {Ramesh}},\ }\bibfield  {title} {\bibinfo {title} {Observation of room-temperature polar skyrmions},\ }\href {https://doi.org/10.1038/s41586-019-1092-8} {\bibfield  {journal} {\bibinfo  {journal} {Nature}\ }\textbf {\bibinfo {volume} {568}},\ \bibinfo {pages} {368}
  (\bibinfo {year} {2019})}\BibitemShut {NoStop}%
\bibitem [{\citenamefont {Luk'yanchuk}\ \emph {et~al.}(2024)\citenamefont {Luk'yanchuk}, \citenamefont {Razumnaya}, \citenamefont {Kondovych}, \citenamefont {Tikhonov},\ and\ \citenamefont {Vinokur}}]{TopoFerroChiral}%
  \BibitemOpen
  \bibfield  {author} {\bibinfo {author} {\bibfnamefont {I.}~\bibnamefont {Luk'yanchuk}}, \bibinfo {author} {\bibfnamefont {A.}~\bibnamefont {Razumnaya}}, \bibinfo {author} {\bibfnamefont {S.}~\bibnamefont {Kondovych}}, \bibinfo {author} {\bibfnamefont {Y.}~\bibnamefont {Tikhonov}},\ and\ \bibinfo {author} {\bibfnamefont {V.~M.}\ \bibnamefont {Vinokur}},\ }\bibfield  {title} {\bibinfo {title} {Topological ferroelectric chirality},\ }\href {https://doi.org/10.48550/arXiv.2406.19728} {\bibfield  {journal} {\bibinfo  {journal} {arXiv:2406.19728}\ } (\bibinfo {year} {2024})}\BibitemShut {NoStop}%
\bibitem [{sup()}]{suppl}%
  \BibitemOpen
  \href@noop {} {}\bibinfo {note} {Supplemental/ancillary file movie1.mp4}\BibitemShut {NoStop}%
\bibitem [{\citenamefont {Chmiel}\ \emph {et~al.}(2018)\citenamefont {Chmiel}, \citenamefont {Waterfield~Price}, \citenamefont {Johnson}, \citenamefont {Lamirand}, \citenamefont {Schad}, \citenamefont {van~der Laan}, \citenamefont {Harris}, \citenamefont {Irwin}, \citenamefont {Rzchowski}, \citenamefont {Eom},\ and\ \citenamefont {Radaelli}}]{Chmiel2018}%
  \BibitemOpen
  \bibfield  {author} {\bibinfo {author} {\bibfnamefont {F.~P.}\ \bibnamefont {Chmiel}}, \bibinfo {author} {\bibfnamefont {N.}~\bibnamefont {Waterfield~Price}}, \bibinfo {author} {\bibfnamefont {R.~D.}\ \bibnamefont {Johnson}}, \bibinfo {author} {\bibfnamefont {A.~D.}\ \bibnamefont {Lamirand}}, \bibinfo {author} {\bibfnamefont {J.}~\bibnamefont {Schad}}, \bibinfo {author} {\bibfnamefont {G.}~\bibnamefont {van~der Laan}}, \bibinfo {author} {\bibfnamefont {D.~T.}\ \bibnamefont {Harris}}, \bibinfo {author} {\bibfnamefont {J.}~\bibnamefont {Irwin}}, \bibinfo {author} {\bibfnamefont {M.~S.}\ \bibnamefont {Rzchowski}}, \bibinfo {author} {\bibfnamefont {C.-B.}\ \bibnamefont {Eom}},\ and\ \bibinfo {author} {\bibfnamefont {P.~G.}\ \bibnamefont {Radaelli}},\ }\bibfield  {title} {\bibinfo {title} {Observation of magnetic vortex pairs at room temperature in a planar $\alpha$-{Fe$_2$O$_3$/Co} heterostructure},\ }\href {https://doi.org/10.1038/s41563-018-0101-x} {\bibfield  {journal} {\bibinfo  {journal} {Nature
  Materials}\ }\textbf {\bibinfo {volume} {17}},\ \bibinfo {pages} {581} (\bibinfo {year} {2018})}\BibitemShut {NoStop}%
\bibitem [{\citenamefont {Jani}\ \emph {et~al.}(2021)\citenamefont {Jani}, \citenamefont {Lin}, \citenamefont {Chen}, \citenamefont {Harrison}, \citenamefont {Maccherozzi}, \citenamefont {Schad}, \citenamefont {Prakash}, \citenamefont {Eom}, \citenamefont {Ariando}, \citenamefont {Venkatesan},\ and\ \citenamefont {Radaelli}}]{Jani2021}%
  \BibitemOpen
  \bibfield  {author} {\bibinfo {author} {\bibfnamefont {H.}~\bibnamefont {Jani}}, \bibinfo {author} {\bibfnamefont {J.-C.}\ \bibnamefont {Lin}}, \bibinfo {author} {\bibfnamefont {J.}~\bibnamefont {Chen}}, \bibinfo {author} {\bibfnamefont {J.}~\bibnamefont {Harrison}}, \bibinfo {author} {\bibfnamefont {F.}~\bibnamefont {Maccherozzi}}, \bibinfo {author} {\bibfnamefont {J.}~\bibnamefont {Schad}}, \bibinfo {author} {\bibfnamefont {S.}~\bibnamefont {Prakash}}, \bibinfo {author} {\bibfnamefont {C.-B.}\ \bibnamefont {Eom}}, \bibinfo {author} {\bibfnamefont {A.}~\bibnamefont {Ariando}}, \bibinfo {author} {\bibfnamefont {T.}~\bibnamefont {Venkatesan}},\ and\ \bibinfo {author} {\bibfnamefont {P.~G.}\ \bibnamefont {Radaelli}},\ }\bibfield  {title} {\bibinfo {title} {Antiferromagnetic half-skyrmions and bimerons at room temperature},\ }\href {https://doi.org/10.1038/s41586-021-03219-6} {\bibfield  {journal} {\bibinfo  {journal} {Nature}\ }\textbf {\bibinfo {volume} {590}},\ \bibinfo {pages} {74} (\bibinfo {year}
  {2021})}\BibitemShut {NoStop}%
\bibitem [{\citenamefont {Amin}\ \emph {et~al.}(2023)\citenamefont {Amin}, \citenamefont {Poole}, \citenamefont {Reimers}, \citenamefont {Barton}, \citenamefont {Dal~Din}, \citenamefont {Maccherozzi}, \citenamefont {Dhesi}, \citenamefont {Nov{\'a}k}, \citenamefont {Krizek}, \citenamefont {Chauhan}, \citenamefont {Campion}, \citenamefont {Rushforth}, \citenamefont {Jungwirth}, \citenamefont {Tretiakov}, \citenamefont {Edmonds},\ and\ \citenamefont {Wadley}}]{Amin2023}%
  \BibitemOpen
  \bibfield  {author} {\bibinfo {author} {\bibfnamefont {O.~J.}\ \bibnamefont {Amin}}, \bibinfo {author} {\bibfnamefont {S.~F.}\ \bibnamefont {Poole}}, \bibinfo {author} {\bibfnamefont {S.}~\bibnamefont {Reimers}}, \bibinfo {author} {\bibfnamefont {L.~X.}\ \bibnamefont {Barton}}, \bibinfo {author} {\bibfnamefont {A.}~\bibnamefont {Dal~Din}}, \bibinfo {author} {\bibfnamefont {F.}~\bibnamefont {Maccherozzi}}, \bibinfo {author} {\bibfnamefont {S.~S.}\ \bibnamefont {Dhesi}}, \bibinfo {author} {\bibfnamefont {V.}~\bibnamefont {Nov{\'a}k}}, \bibinfo {author} {\bibfnamefont {F.}~\bibnamefont {Krizek}}, \bibinfo {author} {\bibfnamefont {J.~S.}\ \bibnamefont {Chauhan}}, \bibinfo {author} {\bibfnamefont {R.~P.}\ \bibnamefont {Campion}}, \bibinfo {author} {\bibfnamefont {A.~W.}\ \bibnamefont {Rushforth}}, \bibinfo {author} {\bibfnamefont {T.}~\bibnamefont {Jungwirth}}, \bibinfo {author} {\bibfnamefont {O.~A.}\ \bibnamefont {Tretiakov}}, \bibinfo {author} {\bibfnamefont {K.~W.}\ \bibnamefont {Edmonds}},\ and\ \bibinfo
  {author} {\bibfnamefont {P.}~\bibnamefont {Wadley}},\ }\bibfield  {title} {\bibinfo {title} {Antiferromagnetic half-skyrmions electrically generated and controlled at room temperature},\ }\href {https://doi.org/10.1038/s41565-023-01386-3} {\bibfield  {journal} {\bibinfo  {journal} {Nature Nanotechnology}\ }\textbf {\bibinfo {volume} {18}},\ \bibinfo {pages} {849} (\bibinfo {year} {2023})}\BibitemShut {NoStop}%
\bibitem [{\citenamefont {Whitehead}(1979)}]{GeorgeWhitehead1979}%
  \BibitemOpen
  \bibfield  {author} {\bibinfo {author} {\bibfnamefont {G.~W.}\ \bibnamefont {Whitehead}},\ }\href {https://doi.org/10.1007/978-1-4612-6318-0} {\emph {\bibinfo {title} {Elements of Homotopy Theory}}},\ Graduate Texts in Mathematics, Vol. 61\ (\bibinfo  {publisher} {Springer New York, NY},\ \bibinfo {year} {1979})\BibitemShut {NoStop}%
\bibitem [{\citenamefont {Climenhaga}\ and\ \citenamefont {Katok}(2017)}]{ClimenhagaKatok}%
  \BibitemOpen
  \bibfield  {author} {\bibinfo {author} {\bibfnamefont {V.}~\bibnamefont {Climenhaga}}\ and\ \bibinfo {author} {\bibfnamefont {A.}~\bibnamefont {Katok}},\ }\href@noop {} {\emph {\bibinfo {title} {From Groups to Geometry and Back}}},\ Student Mathematical Library, Vol. 81\ (\bibinfo  {publisher} {American Mathematical Society},\ \bibinfo {year} {2017})\BibitemShut {NoStop}%
\bibitem [{\citenamefont {Brown}(1982)}]{CohomologyOfGroups}%
  \BibitemOpen
  \bibfield  {author} {\bibinfo {author} {\bibfnamefont {K.~S.}\ \bibnamefont {Brown}},\ }\href {https://doi.org/10.1007/978-1-4684-9327-6} {\emph {\bibinfo {title} {Cohomology of groups}}},\ Graduate Texts in Mathematics, Vol. 87\ (\bibinfo  {publisher} {Springer-Verlag, NY},\ \bibinfo {year} {1982})\BibitemShut {NoStop}%
\bibitem [{\citenamefont {Rotman}(2009)}]{IntroductiontoHomological}%
  \BibitemOpen
  \bibfield  {author} {\bibinfo {author} {\bibfnamefont {J.~J.}\ \bibnamefont {Rotman}},\ }\href {https://doi.org/10.1007/b98977} {\emph {\bibinfo {title} {An Introduction to Homological Algebra}}},\ Universitext\ (\bibinfo  {publisher} {Springer New York, NY},\ \bibinfo {year} {2009})\BibitemShut {NoStop}%
\bibitem [{\citenamefont {Kosniowski}(1980)}]{Kosniowski}%
  \BibitemOpen
  \bibfield  {author} {\bibinfo {author} {\bibfnamefont {C.}~\bibnamefont {Kosniowski}},\ }\href {https://doi.org/10.1017/CBO9780511569296} {\emph {\bibinfo {title} {A First Course in Algebraic Topology}}}\ (\bibinfo  {publisher} {Cambridge University Press},\ \bibinfo {year} {1980})\BibitemShut {NoStop}%
\bibitem [{\citenamefont {Abe}(1940)}]{Abe1940}%
  \BibitemOpen
  \bibfield  {author} {\bibinfo {author} {\bibfnamefont {M.}~\bibnamefont {Abe}},\ }\bibfield  {title} {\bibinfo {title} {Über die stetigen abbildungen der $n$-sphäre in einen metrischen raum},\ }\href {https://doi.org/10.4099/jjm1924.16.0_169} {\bibfield  {journal} {\bibinfo  {journal} {Japanese journal of mathematics: transactions and abstracts}\ }\textbf {\bibinfo {volume} {16}},\ \bibinfo {pages} {169} (\bibinfo {year} {1940})}\BibitemShut {NoStop}%
\bibitem [{\citenamefont {Kobayashi}\ \emph {et~al.}(2012)\citenamefont {Kobayashi}, \citenamefont {Kobayashi}, \citenamefont {Kawaguchi}, \citenamefont {Nitta},\ and\ \citenamefont {Ueda}}]{KOBAYASHI2012577}%
  \BibitemOpen
  \bibfield  {author} {\bibinfo {author} {\bibfnamefont {S.}~\bibnamefont {Kobayashi}}, \bibinfo {author} {\bibfnamefont {M.}~\bibnamefont {Kobayashi}}, \bibinfo {author} {\bibfnamefont {Y.}~\bibnamefont {Kawaguchi}}, \bibinfo {author} {\bibfnamefont {M.}~\bibnamefont {Nitta}},\ and\ \bibinfo {author} {\bibfnamefont {M.}~\bibnamefont {Ueda}},\ }\bibfield  {title} {\bibinfo {title} {Abe homotopy classification of topological excitations under the topological influence of vortices},\ }\href {https://doi.org/10.1016/j.nuclphysb.2011.11.003} {\bibfield  {journal} {\bibinfo  {journal} {Nuclear Physics B}\ }\textbf {\bibinfo {volume} {856}},\ \bibinfo {pages} {577} (\bibinfo {year} {2012})}\BibitemShut {NoStop}%
\bibitem [{\citenamefont {Tiwari}\ and\ \citenamefont {Bzdu\v{s}ek}(2020)}]{PhysRevB.101.195130}%
  \BibitemOpen
  \bibfield  {author} {\bibinfo {author} {\bibfnamefont {A.}~\bibnamefont {Tiwari}}\ and\ \bibinfo {author} {\bibfnamefont {T.}~\bibnamefont {Bzdu\v{s}ek}},\ }\bibfield  {title} {\bibinfo {title} {Non-{Abelian} topology of nodal-line rings in $\mathcal{PT}$-symmetric systems},\ }\href {https://doi.org/10.1103/PhysRevB.101.195130} {\bibfield  {journal} {\bibinfo  {journal} {Phys. Rev. B}\ }\textbf {\bibinfo {volume} {101}},\ \bibinfo {pages} {195130} (\bibinfo {year} {2020})}\BibitemShut {NoStop}%
\bibitem [{\citenamefont {Hilton}(1953)}]{Hilton1953}%
  \BibitemOpen
  \bibfield  {author} {\bibinfo {author} {\bibfnamefont {P.~J.}\ \bibnamefont {Hilton}},\ }\href@noop {} {\emph {\bibinfo {title} {An introduction to homotopy theory}}}\ (\bibinfo  {publisher} {Cambridge University Press},\ \bibinfo {year} {1953})\BibitemShut {NoStop}%
\bibitem [{\citenamefont {Papanicolaou}\ and\ \citenamefont {Tomaras}(1991)}]{PAPANICOLAOU1991425}%
  \BibitemOpen
  \bibfield  {author} {\bibinfo {author} {\bibfnamefont {N.}~\bibnamefont {Papanicolaou}}\ and\ \bibinfo {author} {\bibfnamefont {T.}~\bibnamefont {Tomaras}},\ }\bibfield  {title} {\bibinfo {title} {Dynamics of magnetic vortices},\ }\href {https://doi.org/10.1016/0550-3213(91)90410-Y} {\bibfield  {journal} {\bibinfo  {journal} {Nuclear Physics B}\ }\textbf {\bibinfo {volume} {360}},\ \bibinfo {pages} {425} (\bibinfo {year} {1991})}\BibitemShut {NoStop}%
\bibitem [{\citenamefont {Xu}\ \emph {et~al.}(2023)\citenamefont {Xu}, \citenamefont {Miranda}, \citenamefont {Pereiro}, \citenamefont {Rybakov}, \citenamefont {Thonig}, \citenamefont {Sj\"oqvist}, \citenamefont {Bessarab}, \citenamefont {Bergman}, \citenamefont {Eriksson}, \citenamefont {Herman},\ and\ \citenamefont {Delin}}]{PhysRevResearch.5.043199}%
  \BibitemOpen
  \bibfield  {author} {\bibinfo {author} {\bibfnamefont {Q.}~\bibnamefont {Xu}}, \bibinfo {author} {\bibfnamefont {I.~P.}\ \bibnamefont {Miranda}}, \bibinfo {author} {\bibfnamefont {M.}~\bibnamefont {Pereiro}}, \bibinfo {author} {\bibfnamefont {F.~N.}\ \bibnamefont {Rybakov}}, \bibinfo {author} {\bibfnamefont {D.}~\bibnamefont {Thonig}}, \bibinfo {author} {\bibfnamefont {E.}~\bibnamefont {Sj\"oqvist}}, \bibinfo {author} {\bibfnamefont {P.~F.}\ \bibnamefont {Bessarab}}, \bibinfo {author} {\bibfnamefont {A.}~\bibnamefont {Bergman}}, \bibinfo {author} {\bibfnamefont {O.}~\bibnamefont {Eriksson}}, \bibinfo {author} {\bibfnamefont {P.}~\bibnamefont {Herman}},\ and\ \bibinfo {author} {\bibfnamefont {A.}~\bibnamefont {Delin}},\ }\bibfield  {title} {\bibinfo {title} {Metaheuristic conditional neural network for harvesting skyrmionic metastable states},\ }\href {https://doi.org/10.1103/PhysRevResearch.5.043199} {\bibfield  {journal} {\bibinfo  {journal} {Phys. Rev. Res.}\ }\textbf {\bibinfo {volume} {5}},\ \bibinfo
  {pages} {043199} (\bibinfo {year} {2023})}\BibitemShut {NoStop}%
\bibitem [{\citenamefont {Hoffmann}\ \emph {et~al.}(2017)\citenamefont {Hoffmann}, \citenamefont {Zimmermann}, \citenamefont {M{\"u}ller}, \citenamefont {Sch{\"u}rhoff}, \citenamefont {Kiselev}, \citenamefont {Melcher},\ and\ \citenamefont {Bl{\"u}gel}}]{Hoffmann2017}%
  \BibitemOpen
  \bibfield  {author} {\bibinfo {author} {\bibfnamefont {M.}~\bibnamefont {Hoffmann}}, \bibinfo {author} {\bibfnamefont {B.}~\bibnamefont {Zimmermann}}, \bibinfo {author} {\bibfnamefont {G.~P.}\ \bibnamefont {M{\"u}ller}}, \bibinfo {author} {\bibfnamefont {D.}~\bibnamefont {Sch{\"u}rhoff}}, \bibinfo {author} {\bibfnamefont {N.~S.}\ \bibnamefont {Kiselev}}, \bibinfo {author} {\bibfnamefont {C.}~\bibnamefont {Melcher}},\ and\ \bibinfo {author} {\bibfnamefont {S.}~\bibnamefont {Bl{\"u}gel}},\ }\bibfield  {title} {\bibinfo {title} {Antiskyrmions stabilized at interfaces by anisotropic {Dzyaloshinskii-Moriya} interactions},\ }\href {https://doi.org/10.1038/s41467-017-00313-0} {\bibfield  {journal} {\bibinfo  {journal} {Nature Communications}\ }\textbf {\bibinfo {volume} {8}},\ \bibinfo {pages} {308} (\bibinfo {year} {2017})}\BibitemShut {NoStop}%
\bibitem [{\citenamefont {Nagase}\ \emph {et~al.}(2021)\citenamefont {Nagase}, \citenamefont {So}, \citenamefont {Yasui}, \citenamefont {Ishida}, \citenamefont {Yoshida}, \citenamefont {Tanaka}, \citenamefont {Saitoh}, \citenamefont {Ikarashi}, \citenamefont {Kawaguchi}, \citenamefont {Kuwahara},\ and\ \citenamefont {Nagao}}]{Nagase2021}%
  \BibitemOpen
  \bibfield  {author} {\bibinfo {author} {\bibfnamefont {T.}~\bibnamefont {Nagase}}, \bibinfo {author} {\bibfnamefont {Y.-G.}\ \bibnamefont {So}}, \bibinfo {author} {\bibfnamefont {H.}~\bibnamefont {Yasui}}, \bibinfo {author} {\bibfnamefont {T.}~\bibnamefont {Ishida}}, \bibinfo {author} {\bibfnamefont {H.~K.}\ \bibnamefont {Yoshida}}, \bibinfo {author} {\bibfnamefont {Y.}~\bibnamefont {Tanaka}}, \bibinfo {author} {\bibfnamefont {K.}~\bibnamefont {Saitoh}}, \bibinfo {author} {\bibfnamefont {N.}~\bibnamefont {Ikarashi}}, \bibinfo {author} {\bibfnamefont {Y.}~\bibnamefont {Kawaguchi}}, \bibinfo {author} {\bibfnamefont {M.}~\bibnamefont {Kuwahara}},\ and\ \bibinfo {author} {\bibfnamefont {M.}~\bibnamefont {Nagao}},\ }\bibfield  {title} {\bibinfo {title} {Observation of domain wall bimerons in chiral magnets},\ }\href {https://doi.org/10.1038/s41467-021-23845-y} {\bibfield  {journal} {\bibinfo  {journal} {Nature Communications}\ }\textbf {\bibinfo {volume} {12}},\ \bibinfo {pages} {3490} (\bibinfo {year}
  {2021})}\BibitemShut {NoStop}%
\bibitem [{\citenamefont {Kennedy}(2014)}]{RicardoKennedyThesis2014}%
  \BibitemOpen
  \bibfield  {author} {\bibinfo {author} {\bibfnamefont {R.}~\bibnamefont {Kennedy}},\ }\emph {\bibinfo {title} {Homotopy Theory of Topological Insulators}},\ \href {http://kups.ub.uni-koeln.de/id/eprint/5873} {Ph.D. thesis},\ \bibinfo  {school} {University of Cologne} (\bibinfo {year} {2014})\BibitemShut {NoStop}%
\end{thebibliography}%

\end{document}